\documentclass[universe,article,submit,moreauthors,pdftex]{Definitions/mdpi} 

\firstpage{1} 
\pubvolume{1}
\issuenum{1}
\articlenumber{0}
\pubyear{2021}
\copyrightyear{2020}
\datereceived{} 
\dateaccepted{} 
\datepublished{} 
\hreflink{https://doi.org/} 

\usepackage[normalem]{ulem}  

\Title{Transport coefficients of hyperonic neutron star cores }

\TitleCitation{Transport coefficients of hyperonic neutron star cores }

\Author{Peter Shternin $^{1}$*\orcidA{}, Isaac Vida\~{n}a $^{2}$\orcidB{}}

\AuthorNames{Peter Shternin,  Isaac Vida\~{n}a}

\AuthorCitation{Shternin, P.S; Vida\~{n}a, I.}

\address{%
$^{1}$ \quad Ioffe Insitute, 26 Politekhnicheskaya st., St.\ Petersburg, 194021, Russia\\
$^{2}$ \quad INFN Sezione \ di Catania, Via Santa Sofia 64, 95123 Catania, Italy}

\corres{Correspondence: pshternin@gmail.com}

\abstract{We consider transport properties of the hypernuclear matter in neutron star cores. In particular, we calculate the thermal conductivity, the shear viscosity, and the momentum transfer rates for np$\Sigma^{-}\Lambda e\mu$ composition of dense matter in $\beta$--equilibrium for baryon number densities in the range $0.1-1$~fm$^{-3}$. The calculations are based on baryon interactions treated within the framework of the non-relativistic Brueckner-Hartree-Fock theory. Bare nucleon-nucleon (NN) interactions  are described by the Argonne v18 phenomenological potential supplemented with the Urbana IX three-nucleon force. Nucleon-hyperon (NY) and hyperon-hyperon (YY) interactions are based on the \new{NSC97e and NSC97a models} of the Nijmegen group. We find that the  baryon contribution to transport coefficients is dominated by the neutron one as in the case of neutron star cores \new{containing} only nucleons. In particular, we find that neutrons dominate the total thermal conductivity over the whole range of densities explored and that, due to the onset of $\Sigma^-$ which leads to the deleptonization of the neutron star core, they dominate also the shear viscosity \new{in} the high density region, in contrast with the pure nucleonic case where the lepton contribution is always the dominant one.}


\newcommand{\new}[1]{#1}

\usepackage{soul}

\begin{document}
\nolinenumbers

\section{Introduction}\label{sec:intro}

The interior composition of neutron stars (NSs) -- among the most dense material objects in the Universe -- is not known yet and various phases of dense matter might be expected in their interiors, see, e.g., Ref.~\cite{Blaschke2018ASSL}, for a recent review. With masses of the order of the Solar mass and $\sim 10$~km radii, the density of matter in the innermost regions of NSs, their cores, can reach values few times that of symmetric nuclear matter at saturation, $n_0=0.16$~fm$^{-3}$. In the simplest model, the matter of NS cores consists of uniform fluid of neutrons ($n$), protons ($p$), and leptons (electrons ($e$) and muons ($\mu$)) in equilibrium with respect to the weak interaction ($\beta$-stable matter). However, at densities of about $(2-3)n_0$ simple energy arguments suggest that other baryonic degrees of freedom, such as, for instance, hyperons, can appear. This possibility was first proposed by \citet{Ambartsumyan1960SvA} in 1960 and it has been later extensively studied in great detail by many authors using different phenomenological \cite{BalbergGal97,Balberg99,Glendenning82,Glendenning85,Glendenning87,Glendenning91,Weber89,Knorren95,Schaffner96,Huber98} or microscopical \cite{Schulze95,Schulze98,Baldo98,Vidana00,Baldo00,Vidana00b,Schulze06,Sammarruca09,Dapo10,Schulze11,Lonardoni14,Lonardoni15,Petschauer16,Logoteta19} approaches to the equation of state (EOS) of NS matter with hyperons. However, although the presence of hyperons in NSs seems to be energetically unavoidable, the softening that
their presence induces on the EOS leads to NS maximum masses which are incompatible with the observation of the
unusually high masses of the millisecond pulsars PSR J1903+0327 ($1.667\pm 0.021\,M_\odot$) \cite{Champion08}, PSR J1614$-$2230 ($1.928\pm 0.017\, M_\odot$) \cite{Demorest10}, PSR J0348+0432 ($2.01\pm 0.04\, M_\odot$) \cite{Antoniadis13} and the most recent one PSR J0740+6620 ($2.14^{+0.10}_{-0.09}\, M_\odot$) 
\cite{Cromartie20}. The solution of this problem, commonly known in the literature as the `hyperon puzzle', is not easy and it has become a subject of very active research in the last years, see, e.g., Refs.\ \cite{ChatterjeeVidana16,Vidana2018RSPSA} and references therein for a recent and comprehensive review. \new{An alternative way to circumvent the hyperon puzzle is to invoke the appearance of other hadronic degrees of freedom that push the onset of hyperons to higher densities, such as for instance meson condensates. The possible existence of a Bose--Einstein condensate of negative kaons in the inner core of NSs has been also extensively considered in the literature (see e.g., Refs.\ \cite{Kaplan86,Kaplan86b,Brown94,Thorsson94,Lee96,Glendenning98} and references therein). When the $K^-$ chemical potential, $\mu_{K^-}$, becomes smaller than the one of the electron, the process $e^-\rightarrow K^-+\nu_e$ becomes energetically possible and kaons condensate in the NS interior. The onset density for this process 
was calculated to be in the range $(2.5-5)n_0$ \cite{Thorsson94,Lee96}. However, the appearance of kaon condensation induces also a strong softening of the EOS and consequently leads to the reduction of the NS maximum mass, as in the case of hyperons, to values below the current observational limits. The role of kaons in NSs is out of the scope of the present paper and the interested reader is referred to the original works on this topic \cite{Kaplan86,Kaplan86b,Brown94,Thorsson94,Lee96,Glendenning98} for a comprehensive description of the implications of kaon condensation on the structure and evolution of NSs, and to Ref.\ \cite{Tolos2020PrPNP} for a recent review of the present status of experimental and theoretical developments on kaon physics.
}

Neutron stars are evolving entities, where various dynamical processes are believed to occur. One can mention  among others, for instance, NS cooling, NS oscillations due to undamped instabilities in rotating stars, or magnetic field dissipation. The theoretical modelling of these processes requires knowledge of the transport properties of dense NS matter \cite{Schmitt2018}. In recent studies, the transport properties of ultradense nucleon or pure quark matter have received considerable attention, see e.g., Ref.~\cite{Schmitt2018} for a review. Conversely, the coverage of the strange sector has been much lower. 
Therefore, in this paper \new{we make a  step} in the direction of filling this gap and consider  transport coefficients of the hypernuclear NS core matter. Specifically, we focus on the thermal conductivity and shear viscosity as well as on the momentum transfer rates \new{in binary collisions}.  \new{The latter coefficients describe friction between the components of the matter in the diffusion process. If the diffusion is caused by the external electromagnetic field, these coefficients enter the generalized Ohm law and determine the electrical conductivity,}
 which \new{is} important for studies of the magnetic field evolution of NSs, see, e.g., Refs.\ \cite{Yakovlev1991Ap&SS,Goldreich1992ApJ,Schmitt2018,Dommes2020PhRvD}. 
Our results are based on the classical formalism of the kinetic theory of Fermi systems \cite{BaymPethick} adapted for the NS context, see Refs.~\cite{Schmitt2018,FlowersItoh1979ApJ,Anderson1987,Shternin2013PhRvC,Shternin2020PhRvD}, and references therein. 

The basic problem in the transport theory of strongly interacting systems lies in the poor knowledge of many-body interactions at high densities. Various methods are employed, often having poor control of the underlying assumptions and approximations in the high density region. Even within the same theoretical framework, differences in the microphysical input may \new{lead to} drastically different results. For instance, for nuclear matter it was found that various nuclear potential models, all behaving well at low densities, \new{result in} an order of magnitude difference of the  transport properties at several nuclear densities \cite{Shternin2020PhRvD}. In the case of hypernuclear matter the situation is even more tricky, because the nucleon-hyperon (NY) and hyperon-hyperon (YY) interactions are, unfortunately, still poorly constrained. Contrary to the nucleon-nucleon (NN) interaction, which is fairly well known due to the larger number of existing scattering data and measured properties of nuclei, experimental data in the hyperonic sector (see Refs.\ \cite{ChatterjeeVidana16,Vidana2018RSPSA,Tolos2020PrPNP} and references therein) is not sufficient yet to constrain these interactions with a \new{comparable} precision.

In the present 
work
we calculate the transport coefficients of np$\Sigma^{-}\Lambda e\mu$ matter. The number fractions of  different matter constituents are found from the requirements of electric charge neutrality and $\beta$-stability.
The EOS of the hyperonic matter is obtained within the non-relativistic Brueckner-Hartree-Fock (BHF) approach using realistic NN, NY, and YY interactions, see Ref.\ \cite{Baldo1999Book} for a general overview of this approach and Refs.\ \cite{Schulze95,Schulze98,Baldo98,Baldo00,Vidana00,Vidana00b} for details of its application to hypernuclear matter. \new{As a first step, in this work, we use a couple of interaction models for the NY and YY interactions in order to explore the difference with the results for pure nucleonic compositions; a more systematic study of the dependence of the results on the choice of baryon interaction is left for future works.}
Within the nucleonic sector we use, in particular, the Argonne v18 (Av18 for short) potential \cite{Wiringa1995PhRvC} supplemented with an effective density-dependent two-body force derived from the three-nucleon force of the Urbana IX (UIX for short) model \cite{Carlson1983NuPhA}. For the interactions involving hyperons we employ \new{two versions of} the Nijmegen Soft-Core 97 (NSC97 in the next) meson-exchange model \cite{Stoks1999PhRvC,Rijken1999PhRvC}, namely, \new{the models NSC97e and NSC97a}. Three-body interactions involving hyperons are, unfortunately, still quite uncertain and, therefore, are not included in this work. \new{In the following, since the nucleonic part is described always with the Av18 NN potential plus the UIX NNN force, for simplicity we will use the labels NSC97e and NSC97a when referring to the two sets of models used.}

The paper is organized as follows. In section~\ref{sec:formalism} we briefly review the basic expressions for the transport coefficients of Fermi systems. In section~\ref{sec:bhf} we shortly present the BHF approach of hypernuclear matter. We present the transport coefficients in section~\ref{sec:discuss}, and conclude in section~\ref{sec:conclude}. In what follows,  we set $\hbar=k_B=c=1$.

\section{Transport coefficients}\label{sec:formalism}

Let us briefly review the formalism used to calculate the transport coefficients in the case of neutron star cores with hyperons. A detailed description of this formalism for the nuclear matter case can be found
in Refs.\ \cite{Shternin2013PhRvC,Shternin2017JPhCS, Shternin2020PhRvD}.

The basic expressions for calculating the thermal conductivity $\kappa_c$ and the shear viscosity $\eta_c$ of particle species {\it c}, and the momentum transfer rates $J_{ci}$ in the binary collisions between particle species {\it c} and {\it i} are
\begin{eqnarray}
  \kappa_c &=& \frac{\pi^2}{3} T \frac{n_c}{p_{Fc}} \lambda_c^\kappa, \label{eq:kappa}\\
  \eta_c &=&  \frac{1}{5} n_c p_{Fc} \lambda_c^\eta, \label{eq:eta}\\
  J_{ci}&=&n_c p_{Fc} \left(\lambda_{ci}^D\right)^{-1},\label{eq:Jci}
\end{eqnarray}
where $T$ is the temperature, $n_c$ and $p_{Fc}$ are, respectively, the particle number densities and Fermi momenta, $\lambda^\alpha_c$ with $\alpha=\kappa,\,\eta$, and $\lambda_{ci}^D$ are the effective mean free paths which are, in general, different for the different transport coefficient in question. Having this in mind, below we drop the index $\alpha$ for brevity and reinstall it only when necessary. The effective mean free paths are obtained in the kinetic theory from the solution of the system of linearized transport equations for the distribution functions of the quasiparticles.

The simplest approximation, which however was proved to be very good in degenerate matter, is the lowest-order variational approximation, in which case the system of transport equations reduces to the the system of linear equations for the effective mean free paths
\begin{equation}\label{eq:vareq}
\sum_{i} \Lambda_{ci} \lambda_{i} =1.
\end{equation}
The matrix of this system is the transport matrix $\Lambda_{ci}$, which  is connected to the effective transport cross-sections $\sigma_{ci}$, $\sigma_{ci}'$ of binary collisions as
\begin{eqnarray}
    \Lambda_{cc}&=&\sum_{i} n_i\sigma_{ci} + n_c\sigma_{cc}',\label{eq:transp_diag}\\
    \Lambda_{ci}&=&n_i\sigma'_{ci},\quad i\neq c.\label{eq:transp_nondiag}
\end{eqnarray}
Effective transport cross-sections $\sigma_{ci}$ and $\sigma_{ci}'$ in Equations~(\ref{eq:transp_diag})--(\ref{eq:transp_nondiag}) are obtained by averaging of a squared matrix element of the scattering matrix with certain angular factors. In our case of np$\Sigma^{-}\Lambda e\mu$ matter, particles interact via the electromagnetic and strong forces. Leptons participate only in the electromagnetic interactions, while the strong interaction is the dominant channel for the baryons (see, however, section~\ref{sec:mfp}).
The explicit integral expressions for the transport cross-sections convenient to use for the strong interactions can be found in Ref.~\cite{Shternin2020PhRvD} and the form of these quantities more convenient for electromagnetic interactions is given, e.g.,  in Ref.~\cite{Shternin2018PRD}. 

The momentum transfer rates in Equation~(\ref{eq:Jci}) are usually treated in the lowest-order moment expansion of the kinetic theory which is practically equivalent to the variational method described above. The effective partial mean free paths for binary collisions are
\begin{equation}
    \left(\lambda_{ci}^D\right)^{-1}=n_i\sigma^D_{ci},
\end{equation}
where $\sigma^D_{ci}$ is the transport cross-section for the diffusion problem \cite{Shternin2020PhRvD}.

Notice, that the transport matrices in Equations~(\ref{eq:vareq})--(\ref{eq:transp_nondiag}) are not symmetric. Non-diagonal matrix elements of the transport matrices obey the relations $n_c p_{Fc}^{-1} \Lambda^\kappa_{ci}=n_i p_{Fi}^{-1} \Lambda^\kappa_{ic}$ and $n_c p_{Fc} \Lambda^\eta_{ci}=n_i p_{Fi} \Lambda^\eta_{ic}$ \cite{Shternin2020PhRvD}.
The momentum transfer rates in Equation~(\ref{eq:Jci}) are symmetric, $J_{ci}=J_{ic}$.

In the quasiparticle approximation, only the collisions in the vicinity of the Fermi surface contribute to transport. If the matrix element does not depend on the energy transfer in collisions (which is $\sim T$), one finds the simple scaling rules valid for Fermi systems at low temperatures, $\sigma_{ci}\propto m_c^{*2}m_i^{*2} T^2$, where $m_c^*$ and $m_i^*$ are the quasiparticles effective masses on the Fermi surface \cite{BaymPethick}. This is the case for collisions mediated by the nuclear forces. Transport cross-sections governed by the  electromagnetic interactions in NS cores obey non-Fermi liquid temperature dependence as a result of the dynamical character of the plasma screening in the dominant transverse channel of the electromagnetic interaction \cite{Heiselberg:1993cr,Heiselberg1992NuPhA}. The matrix elements of the `electromagnetic' transport matrix do not follow $\propto T^2$ scaling, but, generally, have weaker temperature dependence; \new{the exact scaling is subject to}  the transport problem in question \cite{Schmitt2018}.

\end{paracol}

\begin{figure}[t]
\widefigure
\begin{minipage}{0.45\textwidth}
\includegraphics[width=0.95\columnwidth]{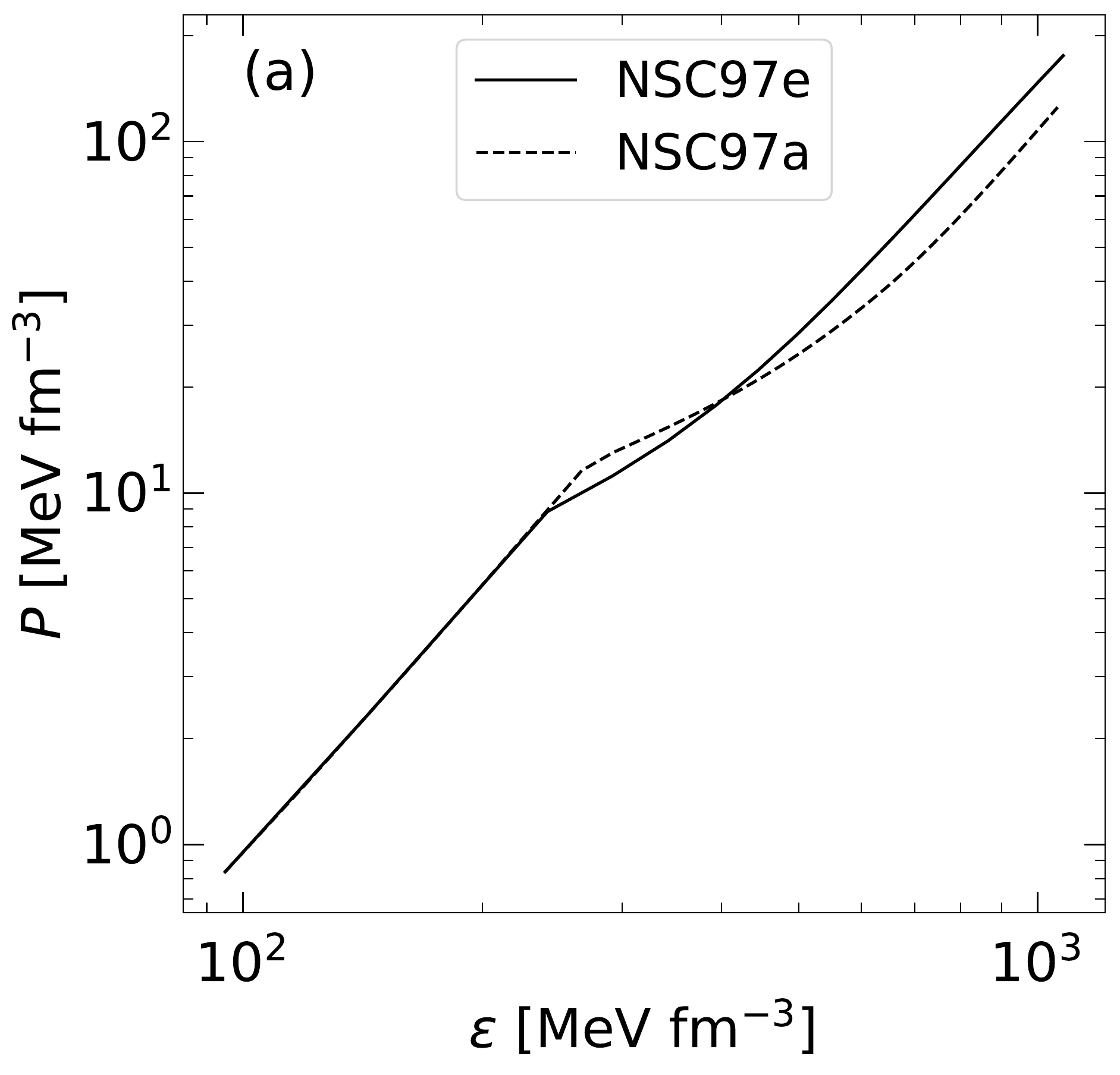}
\end{minipage}
\begin{minipage}{0.45\textwidth}
\includegraphics[width=\columnwidth]{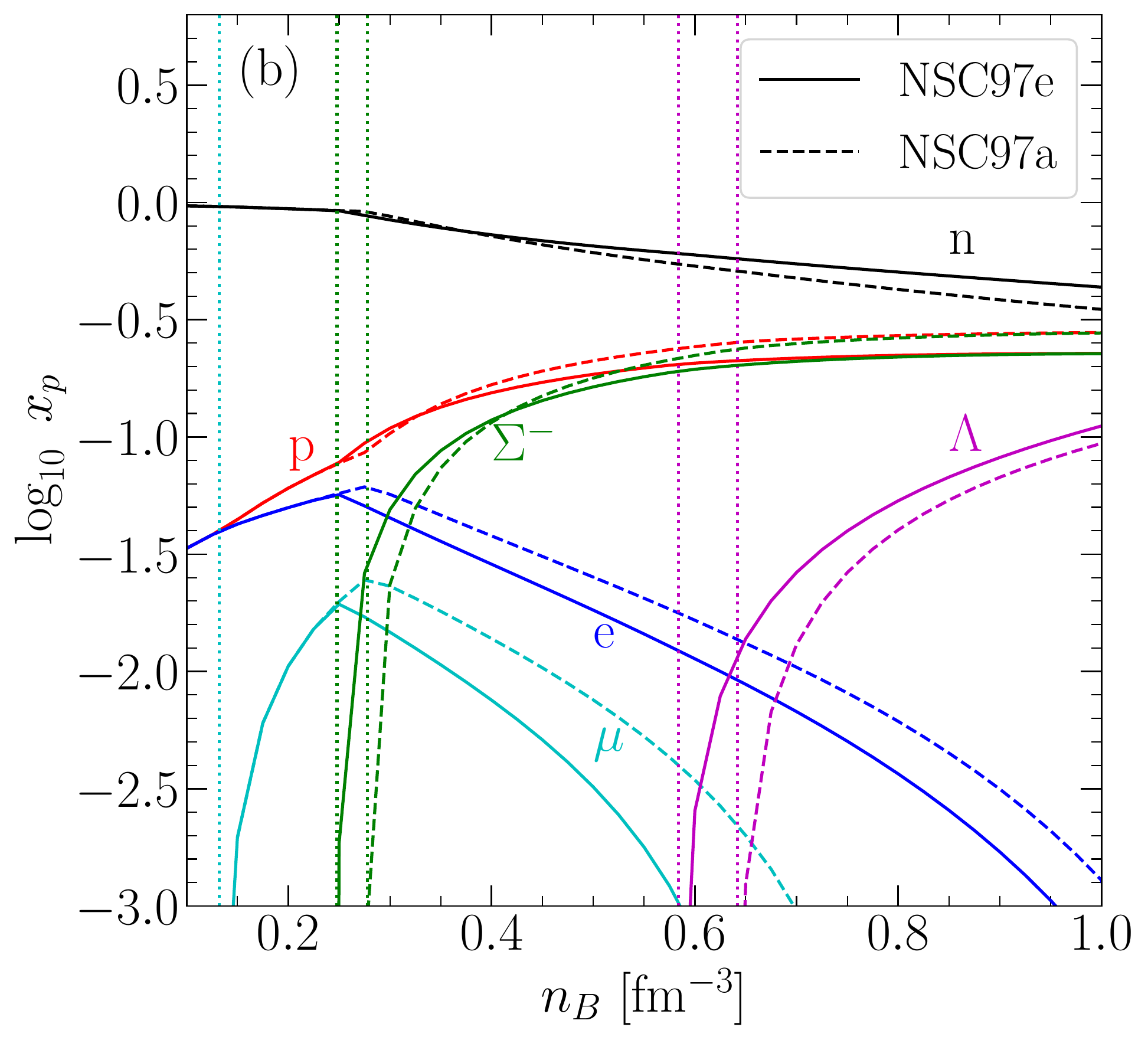}
\end{minipage}
\caption{Properties of $\beta$-stable hypernuclear matter. 
 Panel (\textbf{a}): Pressure as function of a total energy density \new{for the NSC97e (solid lines) and NSC97a (dashed lines) models}. Panel (\textbf{b}): Particle fractions as a function of total baryon number density \new{for the NSC97e (solid lines) and NSC97a (dashed lines) models}. \new{Vertical dashed lines indicate $\mu$, $\Sigma^{-}$, and $\Lambda$ thresholds.}}\label{fig:xs}
\end{figure}  

\begin{paracol}{2}
\switchcolumn

\section{Brueckner--Hartree--Fock approach of hypernuclear matter}\label{sec:bhf}

Calculations of the transport cross-sections, and consequently of the transport coefficients, require as an input the squares of the different scattering transition matrix elements and the effective masses of the involved baryons which should be provided by a nuclear many-body theory. \new{
 Various methods have been considered to solve the nuclear many-body problem: the variational approach \cite{Akmal98}, the correlated basis function (CBF) formalism \cite{Fabrocini93}, the self-consistent Green’s function (SCGF) technique \cite{Kadanoff62,Kraeft86}, or the Brueckner–Bethe–Goldstone (BBG) \cite{Day67} and the Dirac–Bruecker–Hartree–Fock (DBHF) theories \cite{Haar87,Haar87b,Brockman90}. Nevertheless, although all of them have been extensively applied to the study of nuclear matter, up to our knowledge, only the BBG theory in the BHF approximation \cite{ Schulze95,Schulze98,Baldo98,Vidana00,Baldo00,Vidana00b,Schulze06,Schulze11,Petschauer16,Logoteta19}, and very recently the DBHF theory \cite{Sammarruca09}, the V-low-k approach \cite{Dapo10}, and the quantum Monte Carlo method \cite{Lonardoni14,Lonardoni15}
 have been extended to the hyperonic sector.}

In this section we briefly review the non-relativistic Brueckner--Hartree--Fock approximation of hypernuclear matter. \new{We would like to note that, although, this approach is non-relativistic we have found that the speed of sound is, in all the range of densities considered here, always smaller than the speed of light. The reason is that the softening of the EOS induced by the appeareance of the hyperons guarantees causality.} The BHF approximation starts with the construction of all  baryon-baryon (NN, YN, and YY) $G$-matrices which describe the interaction between two baryons in the presence of a surrounding medium. The $G$-matrices are obtained by solving the well-known coupled-channel Bethe--Goldstone integral equation
\begin{eqnarray}
\langle \vec k_{B_1} \vec k_{B_2} |G(\omega)|\vec k_{B_3} \vec k_{B_4} \rangle =
\langle \vec k_{B_1} \vec k_{B_2} |V|\vec k_{B_3} \vec k_{B_4} \rangle\phantom{aaaaaaaa} \nonumber \\ \nonumber \\
+\sum_{B_iB_j}
\frac{\langle \vec k_{B_1} \vec k_{B_2} |V|\vec k_{B_i} \vec k_{B_j} \rangle
 \langle \vec k_{B_i}\vec k_{B_j}|Q|\vec k_{B_i}\vec k_{B_j}\rangle
 \langle \vec k_{B_i} \vec k_{B_j} |G(\omega)|\vec k_{B_3} \vec k_{B_4} \rangle }
 {\omega-E_{B_i}(k_{B_i})-E_{B_j}(k_{B_j})+i\eta}
 \ ,
\label{eq:bbg}
\end{eqnarray}
where $V$ is the bare baryon-baryon interaction, $Q$ is the Pauli operator that prevents the intermediate 
baryons $B_i$ and $B_j$ from being scattered to states below their respective Fermi momenta, $\omega$ is the sum of the non-relativistic single-particle energies of the interacting baryons, and $\eta$ is an infinitesimal positive quantity.

The single-particle energy of a baryon $B_i$ is given by
\begin{equation}
E_{B_i}(k_{B_i})=m_{B_i}+\frac{k^2_{B_i}}{2m_{B_i}}+\mbox{Re}[U_{B_i}(k_{B_i})] \ ,
\label{eq:spe}
\end{equation}
new{where} $m_{B_i}$ denotes the rest mass of the baryon, and the (complex) single-particle potential $U_{B_i}(k_{B_i})$ represents the average field `felt' by the baryon owing to its interaction \new{with  other baryons}. In the BHF approximation, $U_{B_i}(k_{B_i})$ is calculated through the `on-shell' $G$-matrix, and is given by
\begin{equation}
U_{B_i}(k_{B_i})=\sum_{B_j}\sum_{\vec k_{B_j}}n_{B_j}(k_{B_j})
\left\langle \vec k_{B_i} \vec k_{B_j} \left|G\left(\omega=E_{B_i}(\vec k_{B_i})+E_{B_j}(\vec k_{B_j})\right)\right|\vec k_{B_i} \vec k_{B_j}  \right\rangle_{\cal{A}} \ .
\label{eq:spp}    
\end{equation}
Here $n_{B_j}(k_{B_j})$ is the occupation number of the baryon species $B_j$, and the index $\cal{A}$ indicates that the matrix elements are properly antisymmetrized when baryons $B_i$ and $B_j$ belong to the same isomultiplet. We note here that the so-called continuous prescription has been adopted for the single-particle potentials when solving the Bethe--Goldstone equation, since, as it was shown in Refs.\ \cite{Song98,Song00}, the contribution to the energy per particle from three-hole line diagrams is minimized in this prescription. As we already said in the introduction all calculations have been carried out with the Av18 NN potential \cite{Wiringa1995PhRvC} supplemented with the UIX three-nucleon force \cite{Carlson1983NuPhA}, which, for the use in the BHF calculations, was reduced to a two-body density-dependent force by averaging over the spatial, spin, and isospin coordinates of the third nucleon in the medium \cite{Loiseau71, Grange76,BaldoFerreira99}. This three-nucleon force contains two parameters that are fixed by requiring that the BHF calculation reproduces the energy and saturation density of the symmetric nuclear matter. For the NY and YY interactions, we have employed the NSC97e and NSC97a models \cite{Stoks1999PhRvC,Rijken1999PhRvC}. \new{The reason behind the choice of these models is} 
that both models result in the best predictions for the hypernuclear observables
among the \new{potentials constructed by} the Nijmegen group. Three-body forces involving hyperons, {\it i.e.,} forces of the type NNY, NYY, and YYY, have been ignored in the calculation due to the large uncertainties still existing about these kind of forces. 

\begin{figure}[h]
\begin{center}
\includegraphics[width=8  cm]{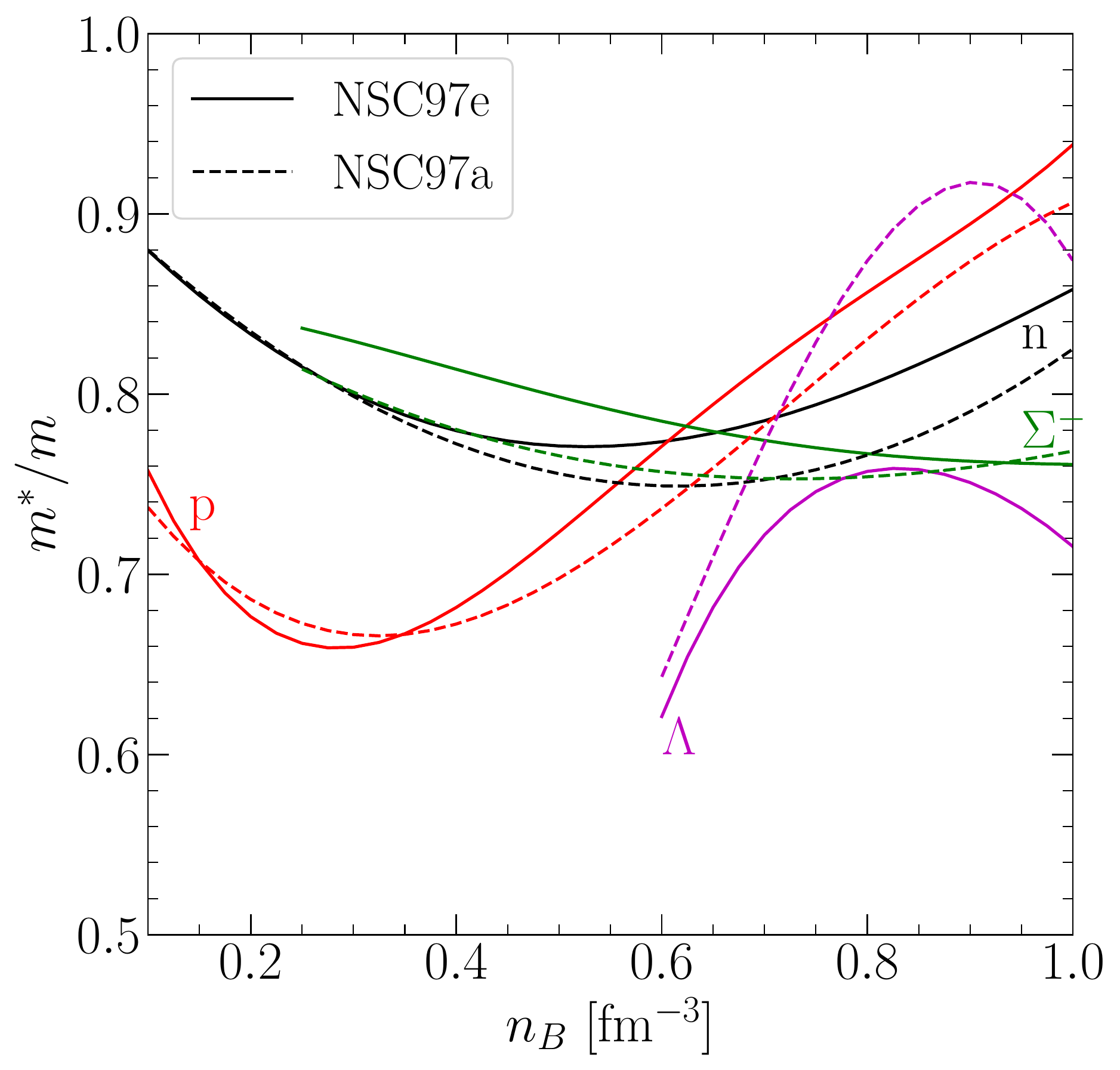}
\end{center}
\caption{Effective baryon masses on the Fermi surface as a function of the total baryon number density \new{for the NSC97e (solid lines) and NSC97a (dashed lines) models}.}\label{fig:meff}
\end{figure}

\new{In order to solve Equations~(\ref{eq:bbg})--(\ref{eq:spp}), one starts from a reasonable guess for the single-particle potentials and obtain initial value of the $G$-matrix from \ Equation~(\ref{eq:bbg}), which is solved in a partial wave basis including contributions up to a total angular momentum $J=4$. Then the new single-particle potential is computed from Equation~(\ref{eq:spp}) and passed as an input to the Equations~(\ref{eq:bbg})--(\ref{eq:spe}). These iterations continue until the desired level of convergence  is reached}. Once a self-consistent solution of Equations\ (\ref{eq:bbg})--(\ref{eq:spp}) is obtained, the baryon contribution to the total energy density can be calculated simply as  
\begin{equation}
\varepsilon_B=n_B\sum_{B_i}\sum_{\vec k_{B_i}}n(k_{B_i})\left(m_{B_i}+\frac{k^2_{B_i}}{2m_{B_i}}+\frac{1}{2}\mbox{Re}[U_{B_i}(k_{B_i}) ] \right)\ , 
\label{eq:ea}    
\end{equation}
where $n_B$ is the baryon number density. Adding to $\varepsilon_B$ the contribution \new{from the} noninteracting leptons, $\varepsilon_L$, the composition and the EOS of neutron star matter can then be obtained from the requirement of equilibrium under weak interaction processes, $\mu_i=b_i\mu_n-q_i\mu_e$ ($b_i$ and $q_i$ denoting the baryon number and electric charge of the species $i$, respectively), and electric charge neutrality, $\sum_iq_in_i=0$. The chemical potentials of  various species and the pressure are computed from the usual thermodynamic relations, $\mu_i=\partial\varepsilon/\partial n_i$ and $P=n_B\partial\varepsilon/\partial n_B-\varepsilon$, where  $\varepsilon=\varepsilon_B+\varepsilon_L$ is the total energy density. The EOS (pressure versus total energy density) and the particle fractions $x_i=n_i/n_B$ of the different species are shown respectively in panels (a) and (b) of Figure~\ref{fig:xs}. \new{Vertical dashed lines in Figure~\ref{fig:xs}b show the threshold densities for appearance of muons ($0.132$~fm$^{-3}$), $\Sigma^{-}$-hyperons ($0.248$~fm$^{-3}$ for NSC97e and $0.278$~fm$^{-3}$ for NSC97a), and $\Lambda$-hyperons ($0.584$~fm$^{-3}$ for NSC97e and $0.642$~fm$^{-3}$ for NSC97a).}

The $G$-matrices needed in the calculation of the transport-cross sections are taken on-shell and on the corresponding Fermi surfaces of the involved baryons, {\it i.e.,} the starting energy is taken as $\omega=E_{B_i}(k_{F_{B_i}})+E_{B_j}(k_{F_{B_j}})$. The effective masses of the different baryons, required in the calculation of the transport coefficients, are calculated from their corresponding single-particle energies as
\begin{equation}\label{eq:meff}
\frac{m^*_{B_i}(k_{B_i}) }{m_{B_i}}=\frac{k_{B_i}}{m_{B_i}}\left(\frac{dE_{B_i}(k_{B_i})}{dk_{B_i}}\right)^{-1} \ .  \end{equation}
Figure~\ref{fig:meff} shows the effective masses of the different baryons evaluated at each density at their corresponding Fermi momenta. \new{The calculation of the effective masses via Equation~(\ref{eq:meff}) requires numerical differentiation of the single-particle potentials resulting from the solution of the BBG equations, which is a delicate numerical task. The results frequently contain considerable numerical noise. It is customary to approximate the results with some smooth function of the baryon density \cite{Baldo2014PhRvC}. Here we employ this approach and fit $m^*_c(n_B)$ functions with fourth-order polynomials. The fit results are given in Appendix~\ref{app:meff}}


\section{Results and Discussion}\label{sec:discuss}

\subsection{Transport matrices and mean free paths}\label{sec:mfp}

\end{paracol}
\begin{figure}[H]
\widefigure
\begin{minipage}{0.45\textwidth}
\includegraphics[width=\columnwidth]{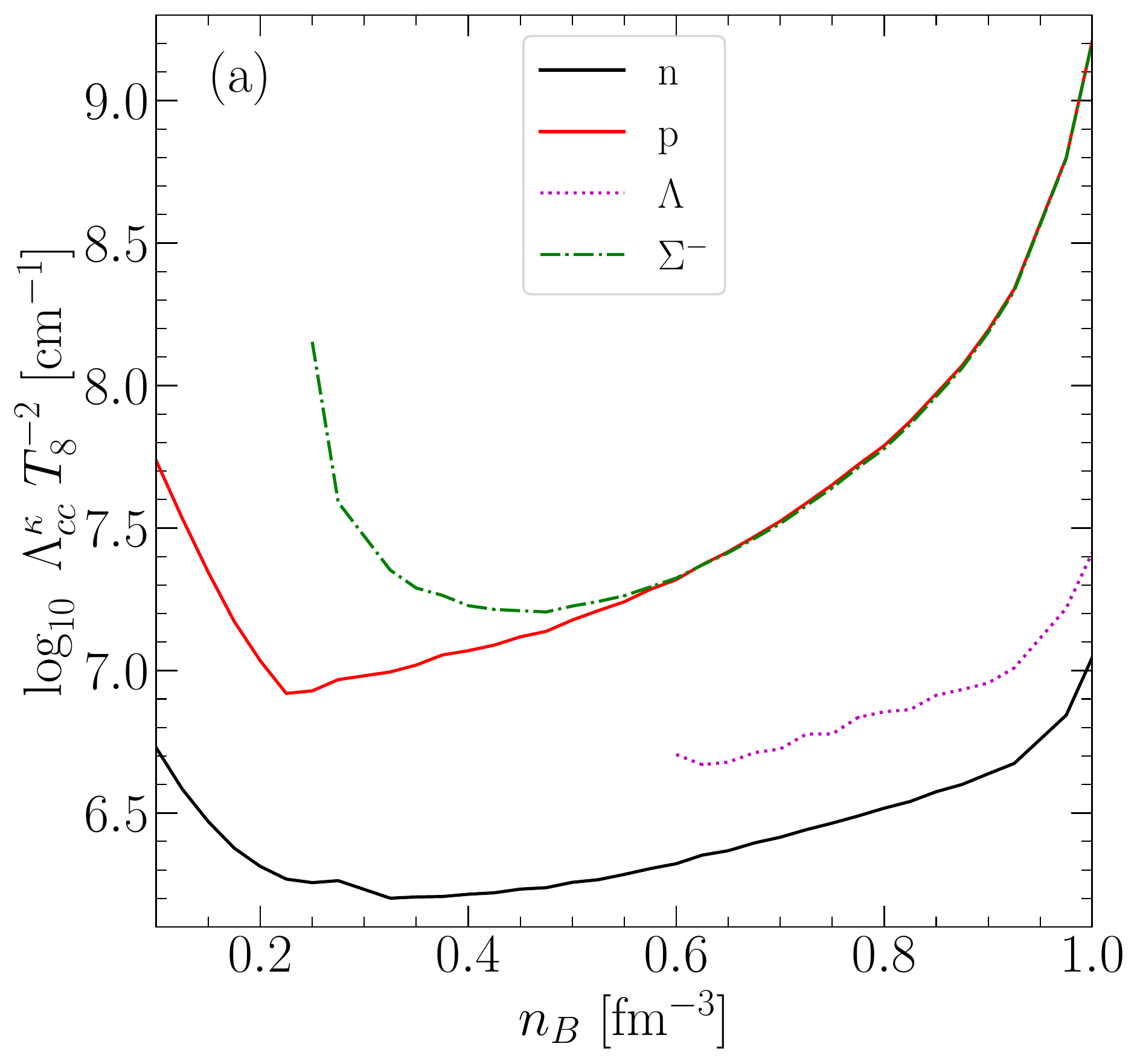}
\end{minipage}
\begin{minipage}{0.45\textwidth}
\includegraphics[width=\columnwidth]{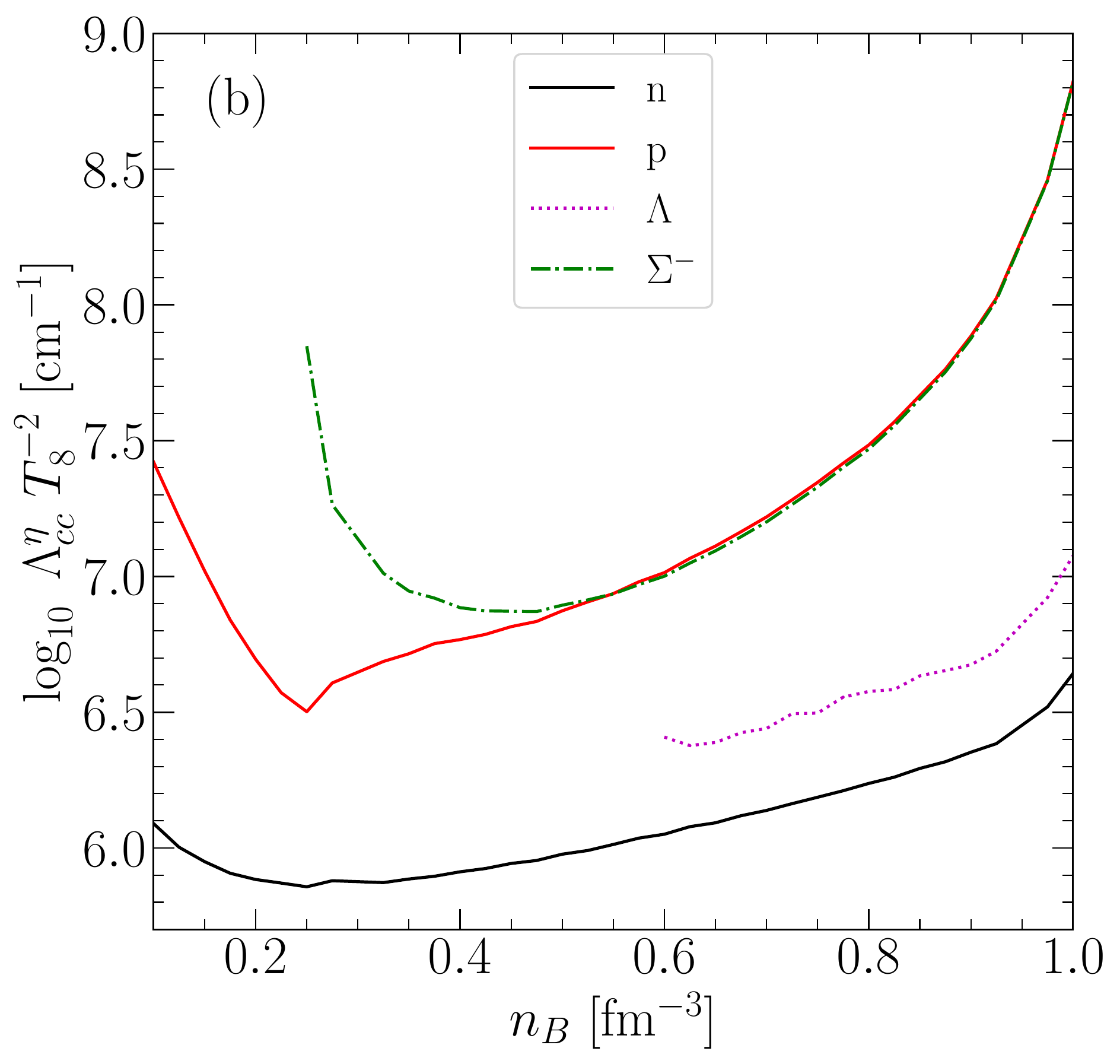}
\end{minipage}
\caption{Diagonal elements of the transport matrix for baryons as a function of the total baryon  number density. Different line types correspond to different particles, as indicated in the legend. Only the strong interaction is taken into account in this plot and temperature-independent ($\Lambda_{cc}T_8^{-2}$) combinations are shown. Results are shown \new{for the NSC97e model}, respectively, for the thermal conductivity (panel \textbf{(a)}) and the shear viscosity cases (panel  \textbf{(b)}).} \label{fig:lambda_cc}
\end{figure}
\begin{paracol}{2}
\switchcolumn

\new{We have found that the results for the NSC97e and NSC97a models are qualitatively similar. Therefore for brevity we mainly focus on the NSC97e model until indicated otherwise.}

The transport matrix $\Lambda_{ci}$ in Equation~(\ref{eq:vareq}) contains a diagonal part [Equation\ (\ref{eq:transp_diag})] and non-diagonal elements [Equation\ (\ref{eq:transp_nondiag})]. If the non-diagonal elements can be neglected (as it is frequently the case, see below), the equations for the different species decouple and the effective mean free paths are given by a simple formula $\lambda_{c}=\Lambda_{cc}^{-1}$.
The diagonal elements of the transport matrix thus represent useful approximations for the inverse mean free paths. In Figure~\ref{fig:lambda_cc} 
we show the diagonal components of the transport matrix for the baryon sector for thermal conductivity [panel (a)] and shear viscosity cases [panel (b)]. In this figure only the strong interaction channels are included and temperature-independent combinations $\Lambda_{cc}T_8^{-2}$ are shown, with $T_8=T/(10^8\,\mbox{K})$. The behavior of $\Lambda^\kappa_{cc}$ and $\Lambda^\eta_{cc}$ is qualitatively similar. The lowest values of $\Lambda_{cc}$ are obtained for neutrons. Thus the neutrons are expected to have the largest mean free paths and dominate the baryon transport. The protons and $\Sigma^{-}$ hyperons scatter one-two orders of magnitude more effectively, and the values of $\Lambda_{cc}$ for the $\Lambda$ hyperon lie in between. Notice  almost identical results for protons and $\Sigma^{-}$ hyperons in Figure~\ref{fig:lambda_cc} at $n_B\gtrsim 0.5$~fm$^{-3}$. 

\end{paracol}
\begin{figure}[t]
\widefigure
\includegraphics[width=12.5 cm]{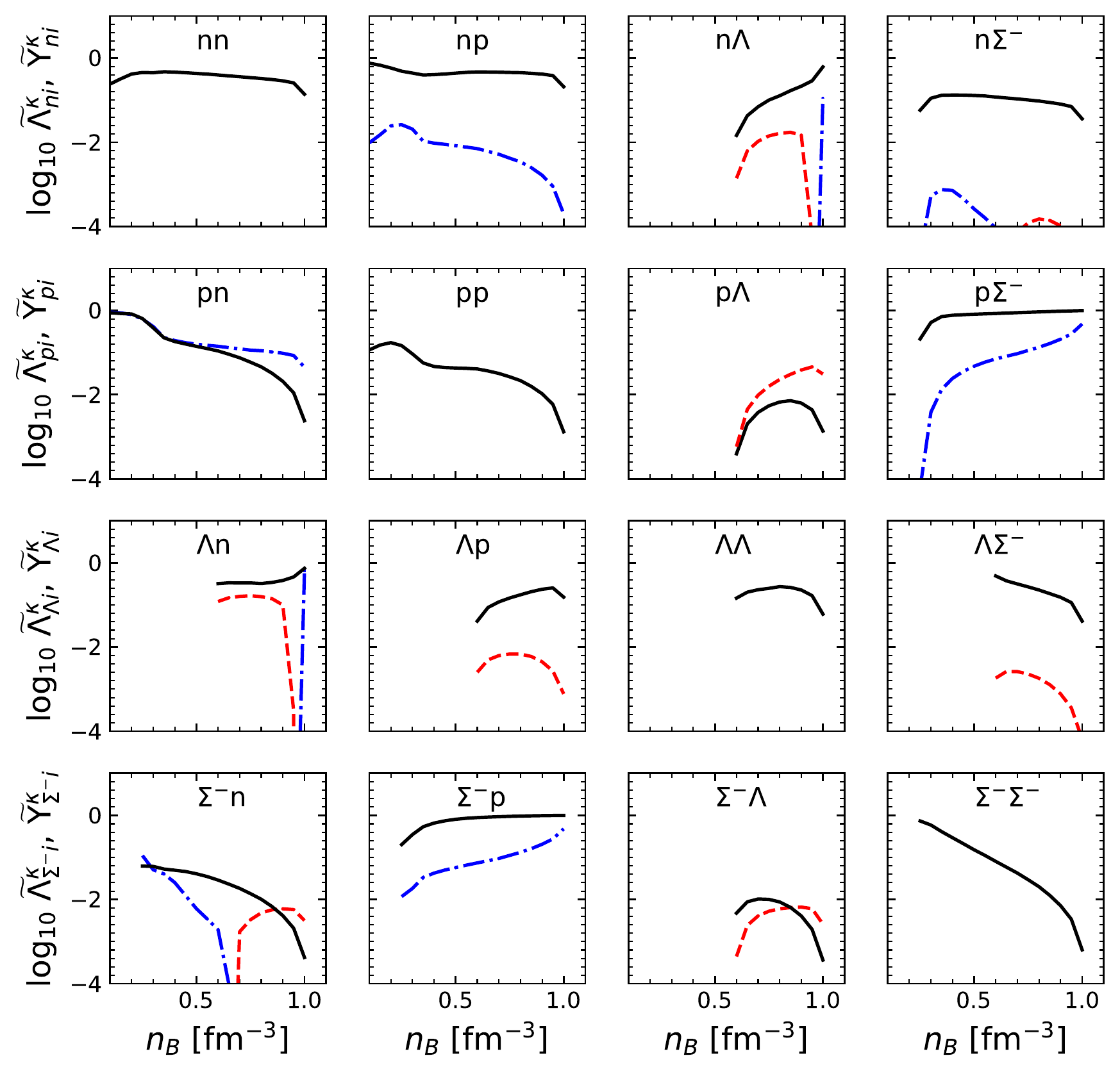}
\caption{Normalized elements of the transport matrix for baryons for the thermal conductivity case \new{and the NSC97e model}. Solid lines in each row \new{($c$=$n$, $p$, $\Lambda$, $\Sigma^-$)} show relative contributions $\widetilde{\Upsilon}^\kappa_{ci}$ to diagonal transport matrix elements from collisions with different baryon species. Other lines show 
the non-diagonal relative matrix elements $\widetilde{\Lambda}^\kappa_{ci}$, where red dashed lines correspond to positive values and the blue dash-dotted ones to negative values, respectively. Only strong forces are included, see text for details.}\label{fig:lambdakappa}
\end{figure}
\begin{paracol}{2}
\switchcolumn

\end{paracol}

\begin{figure}[t]
\widefigure
\includegraphics[width=12.5 cm]{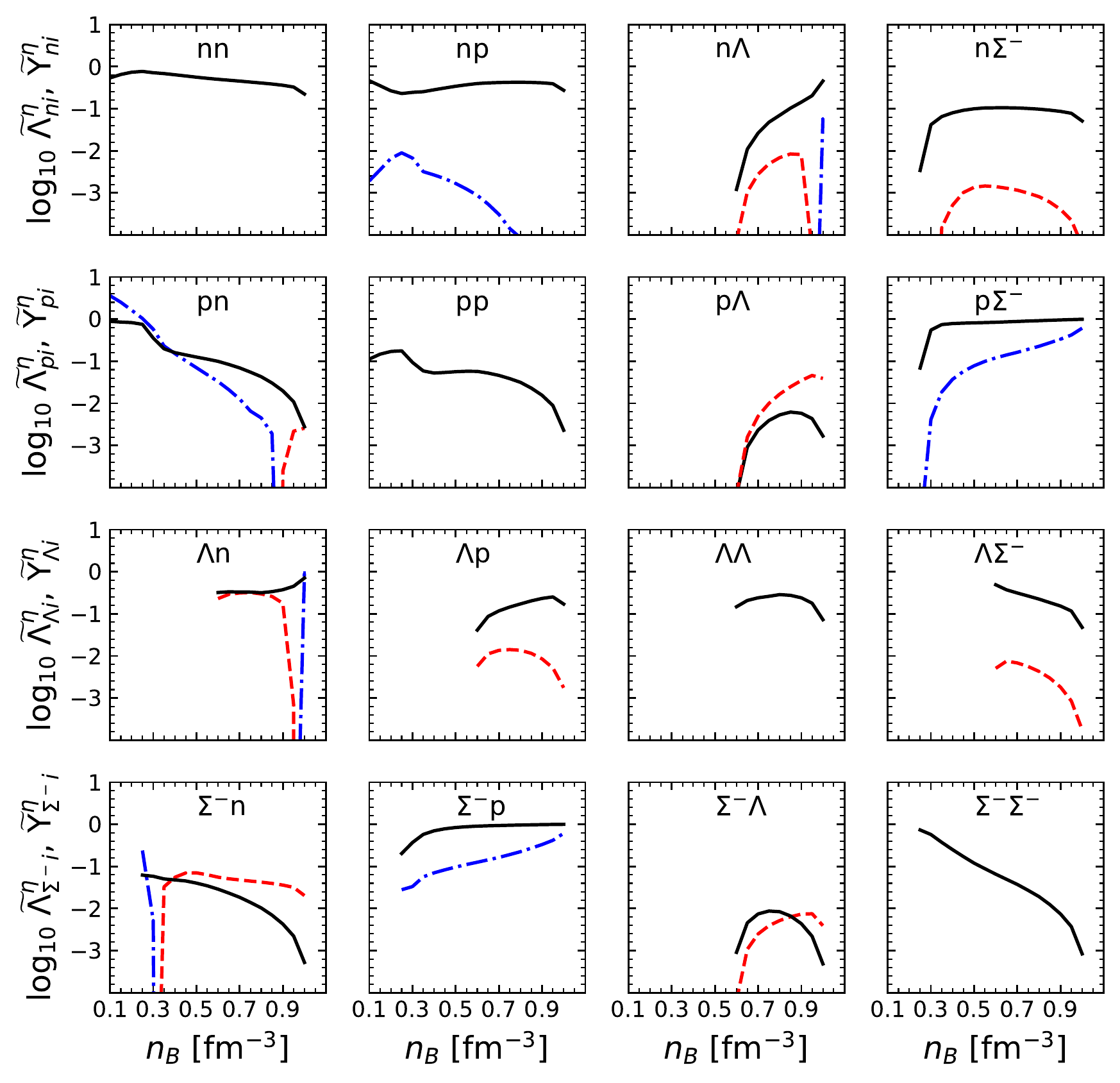}
\caption{Normalized elements of the transport matrix \new{for} baryons for the shear viscosity case \new{and NSC97e model}. Solid lines in each row \new{($c$=$n$, $p$, $\Lambda$, $\Sigma^-$)} show relative contributions $\widetilde{\Upsilon}^\eta_{ci}$ to diagonal transport matrix elements. Other lines show 
the non-diagonal relative matrix elements $\widetilde{\Lambda}^\eta_{ci}$, where red dashed lines correspond to positive values and the blue dash-dotted ones to negative values, respectively. Only strong forces are included, see text for details. }\label{fig:lambdaeta}
\end{figure}
\begin{paracol}{2}
\switchcolumn

\new{These results} can be understood by analyzing the partial contributions from the scattering on different species in Equation~(\ref{eq:transp_diag}). In the matrix-shaped Figures~\ref{fig:lambdakappa} and \ref{fig:lambdaeta} with the solid lines we plot the relative contributions to $\Lambda_{cc}$ for thermal conductivity and shear viscosity cases, respectively. The rows in these figures are numerated with index $c$ and columns with index $i$ of Equation~(\ref{eq:transp_diag}). Namely, we show in each panel with solid lines the ratio $\widetilde{\Upsilon}_{ci}=(n_i\sigma_{ci}+\delta_{ci}n_c\sigma_{cc}')\Lambda_{cc}^{-1}$ \new{(notice that the collisions of the like species contain additional $n_c\sigma'_{cc}$ contribution)}. It is clear that soon after the $\Sigma^{-}$ appearance, their mutual \new{collisions} with protons become completely dominant scattering mechanisms of these species. Taking into account that at $n_B\gtrsim 0.5$~fm$^{-3}$ the proton and $\Sigma^{-}$ number fractions are similar due to the requirement of charge neutrality in lack of charged leptons, see Figure~\ref{fig:xs}b, the coincidence of their transport matrix elements in Figure~\ref{fig:lambda_cc} 
is natural. Because of the large values of the transport matrix elements, the proton and $\Sigma^-$ mean free paths, $\lambda_p$ and $\lambda_{\Sigma^{-}}$, are small and, therefore, these two baryons are not expected to influence the transport equations for other species. Having this in mind and \new{taking into account} the relatively small fraction of these particles (in comparison to neutrons) in $\beta$-stable matter, one expects that they do not give sizable contribution to overall transport coefficients.


In order to analyze the mutual impact of the mean free paths of different species, it is instructive to introduce the renormalized  mean free paths defined as $\widetilde{\lambda}_c = \lambda_c\Lambda_{cc}$. Then the system  (\ref{eq:vareq}) takes the form
\begin{equation}\label{eq:vareq_renorm}
    1=\sum_{i}\frac{{\Lambda}_{ci}}{\Lambda_{ii}}\widetilde{\lambda}_i \equiv\sum_{i}{\widetilde{\Lambda}}_{ci}\widetilde{\lambda}_i,
\end{equation}
where the renormalized transport matrix elements $\widetilde{\Lambda}_{ci}=\Lambda_{ci}/\Lambda_{ii}$ are introduced.

With red dashed (for positive values) and blue dash-dotted (for negative values) lines in Figures~\ref{fig:lambdakappa} and \ref{fig:lambdaeta} we show logarithms of the non-diagonal relative matrix elements $\log_{10} |\widetilde{\Lambda}_{ci}|$, $c\neq i$. In the majority of cases, $|\widetilde{\Lambda}_{ci}|\ll1$, which means that the influence of $\lambda_i$ on the equations for the $c$ species in that cases is negligible. Notice that, irrespectively of the  values of $|\widetilde{\Lambda}_{ci}|$ where one of the indices refers to  protons or $\Sigma^{-}$ hyperons, the corresponding terms can be safely neglected since  $\lambda_p$ and $\lambda_{\Sigma^{-}}$ are small as discussed above. This means that protons and  $\Sigma^{-}$ hyperons can be treated as passive scatterers. 
The remaining $\widetilde{\Lambda}_{ci}$ matrix elements correspond to the $n\Lambda$ subsystem. One observes that $\widetilde{\Lambda}^\eta_{\Lambda n}$ and $\widetilde{\Lambda}^\eta_{n\Lambda}$  and also, for the thermal conductivity case and at the largest densities, $\widetilde{\Lambda}^\kappa_{\Lambda n }$ and $\widetilde{\Lambda}^\kappa_{n\Lambda }$, are significant  and may need to be taken into account. 

Up to now we have discussed the collisions in the baryon subsystem mediated by the strong interaction. The charged particles, $p\Sigma^{-}e\mu$ in our case, \new{also} participate in the electromagnetic interactions\footnote{Neutrons and $\Lambda$ hyperons also couple to the electromagnetic fields due to their magnetic moments, however, this contribution is negligible.}. We neglect the interference between the electromagnetic and strong amplitudes and consider these interaction channels separately. The important difference between the mean free paths governed by electromagnetic and strong interactions in NS cores is in the non-Fermi-liquid temperature dependence of the former. This \new{stems from} the dynamical character of the plasma screening in the dominant transverse channel of the electromagnetic interaction \cite{Heiselberg:1993cr,Heiselberg1992NuPhA}. \new{As already stated, the} matrix elements of the `electromagnetic' transport matrix do not follow the \new{the same} $\propto T^2$ scaling \new{as the matrix elements for the strong interaction}.
For shear viscosity the contribution of the electromagnetic interactions of baryons is several orders of magnitude smaller than the contribution of strong interactions and, therefore, it can be always neglected, see e.g., Ref.\ \cite{Shternin2020PhRvD}. For the thermal conductivity case at low temperatures, the diagonal elements of the transport matrix of charged particles are found to be given by a simple universal expression
\begin{equation}
\Lambda^{\kappa,em}_{cc}
=\frac{6\zeta(3)}{\pi^2} \alpha_f T=2.45\times 10^6\, T_8\ \mathrm{cm}^{-1},
\end{equation}
irrespectively of the content of the charged particles in the matter. For protons and $\Sigma^{-}$ hyperons at sufficiently low temperatures (e.g., $T\lesssim 10^7$~K) it can become comparable with $\Lambda^\kappa_{cc}$ \new{resulting} from strong interactions. This makes the mean free paths of $p$ and $\Sigma^{-}$ even smaller \new{so that} they can be obtained by treating the baryon subsystem separately.
Furthermore, non-diagonal matrix elements of $\Lambda_{ci}^{em}$ are small both for viscosity and thermal conductivity \cite{Shternin2020PhRvD}. As a consequence, lepton and baryon subsystems can indeed be treated separately. Moreover, in both subsystems $p$ and $\Sigma^{-}$ act as passive scatterers. The influence of their non-equilibrium distributions on transport properties of other constituents is negligible. The combined contribution of $p$  and $\Sigma^{-}$ to the total values of transport coefficients is found to be less than 10\% for thermal conductivity and less then 5\% for shear viscosity. 
In what follows, the transport coefficients of leptons are calculated according to Refs.~\cite{ShterninYakovlev2007,ShterninYakovlev2008}.

\subsection{Corrections to the variational solution}\label{sec:exact}
The variational solution gives a first approximation to the transport coefficients.
In a multicomponent Fermi-liquid it is possible to obtain the exact solutions of the corresponding system of transport equations \cite{FlowersItoh1979ApJ,Anderson1987}. However, because of the non-Fermi-liquid behavior of the matrix elements of the electromagnetic interaction, obtaining the exact solution for the full $np\Sigma^{-}\Lambda e\mu$ mixture is complicated and standard approaches \cite{FlowersItoh1979ApJ,Anderson1987} do not work. Nevertheless, as discussed above, it is sufficient to consider the  $n\Lambda$ and $e\mu$ subsystems separately, with protons and $\Sigma^{-}$ particles acting as the passive scatterers. The corrections to the variational solution in the electromagnetic sector were considered in Refs.~\cite{ShterninYakovlev2007,ShterninYakovlev2008} and were found to be less than 5-10\%. 

\begin{figure}[H]
\begin{center}
\includegraphics[width=8cm]{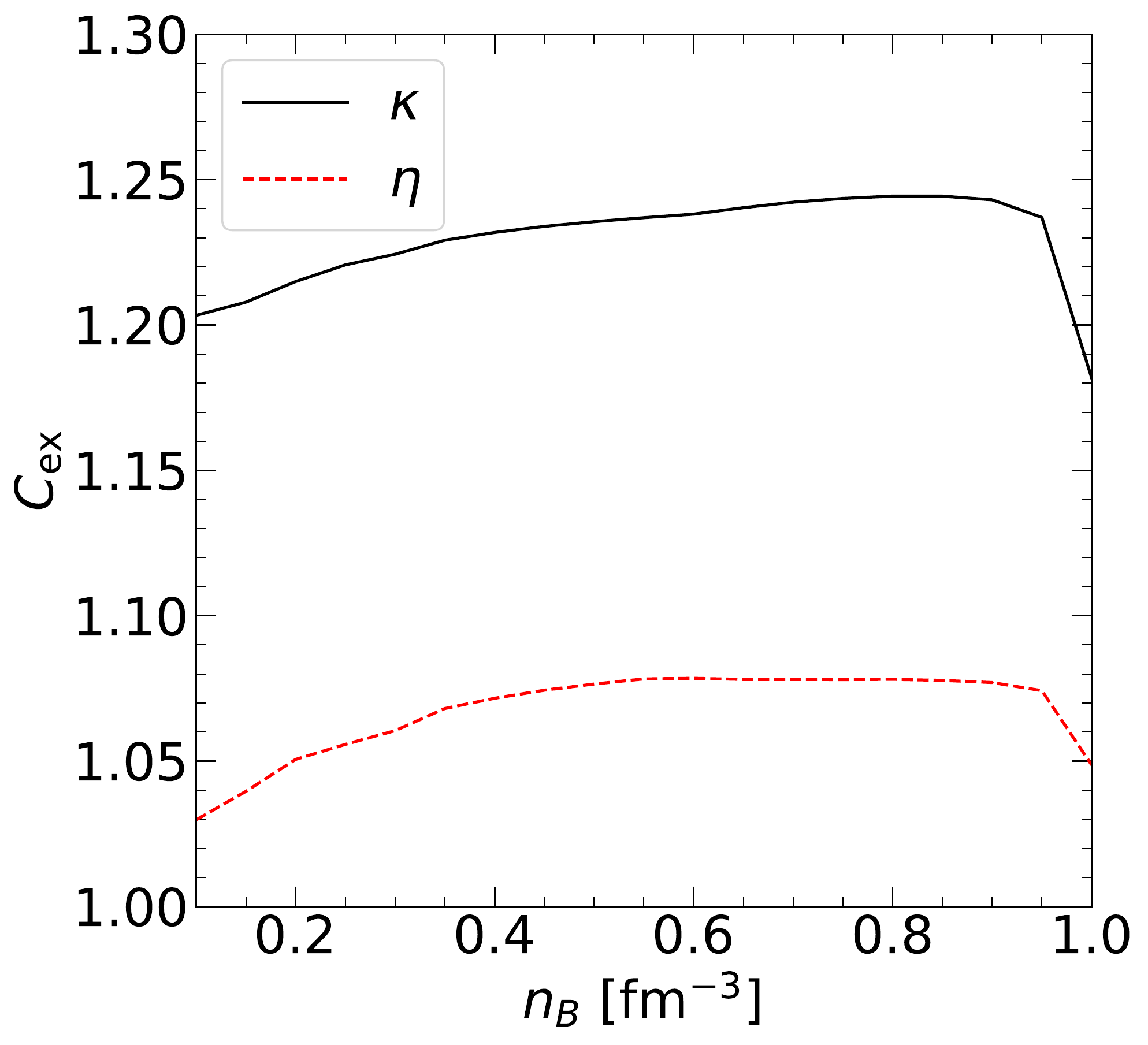}
\end{center}
\caption{Corrections to the variational solution for the $n\Lambda$ subsystem \new{(obtained with the NSC97e model)}.
Solid and dashed lines show correction coefficients for the thermal conductivity and shear viscosity cases, respectively.}
\label{fig:Cex}
\end{figure}

The corrections to the variational solution in the $n\Lambda$ subsystem can be calculated using the method described in Ref.~\cite{Anderson1987}, or by the numerical solution of the system of integral equations \cite{Shternin2017JPhCS}. It is customary to express transport coefficients resulting from the exact solution, via the correction coefficients
\begin{equation}
    \kappa_\mathrm{ex}=C^\kappa_\mathrm{ex} \kappa_\mathrm{var},\quad 
    \eta_\mathrm{ex}=C^\eta_\mathrm{ex} \eta_\mathrm{var}.
\end{equation}
In Figure~\ref{fig:Cex} we show the correction factors $C_\mathrm{ex}$ for the thermal conductivity (solid line) and shear viscosity (dashed line) for the $n\Lambda$ subsystem, where protons and $\Lambda$-hyperons are treated as passive scatterers. The results are similar as for the nuclear matter \cite{Shternin2013PhRvC,Shternin2017JPhCS,Shternin2020PhRvD}. For the shear viscosity, the correction coefficient does not exceed 5\%; this correction is truly minor having in mind other systematic uncertainties and it can be neglected. The correction coefficient for the thermal conductivity case is about 20\% for the $n\Lambda$ and can be included in calculations. 

We do not discuss the corrections to the simplest approximation for the momentum transfer rates $J_{ci}$. These quantities are required as an input to the studies of the magnetic field evolution in NS cores. In this case the exact solution of the transport equations in presence of the magnetic field is required. Such solutions have never been constructed for multicomponent Fermi-systems to the best of our knowledge and require a separate study. Since similar transport cross-sections (with the same leading-order dependencies) enter expressions for the shear viscosity and the momentum  transfer rates, one can expect that the corrections resulting from better approximations to the transport equations will not be large. 


\subsection{Transport coefficients}

\end{paracol}
\begin{figure}[H]
\widefigure
\begin{minipage}{0.32\textwidth}
\includegraphics[width=\textwidth]{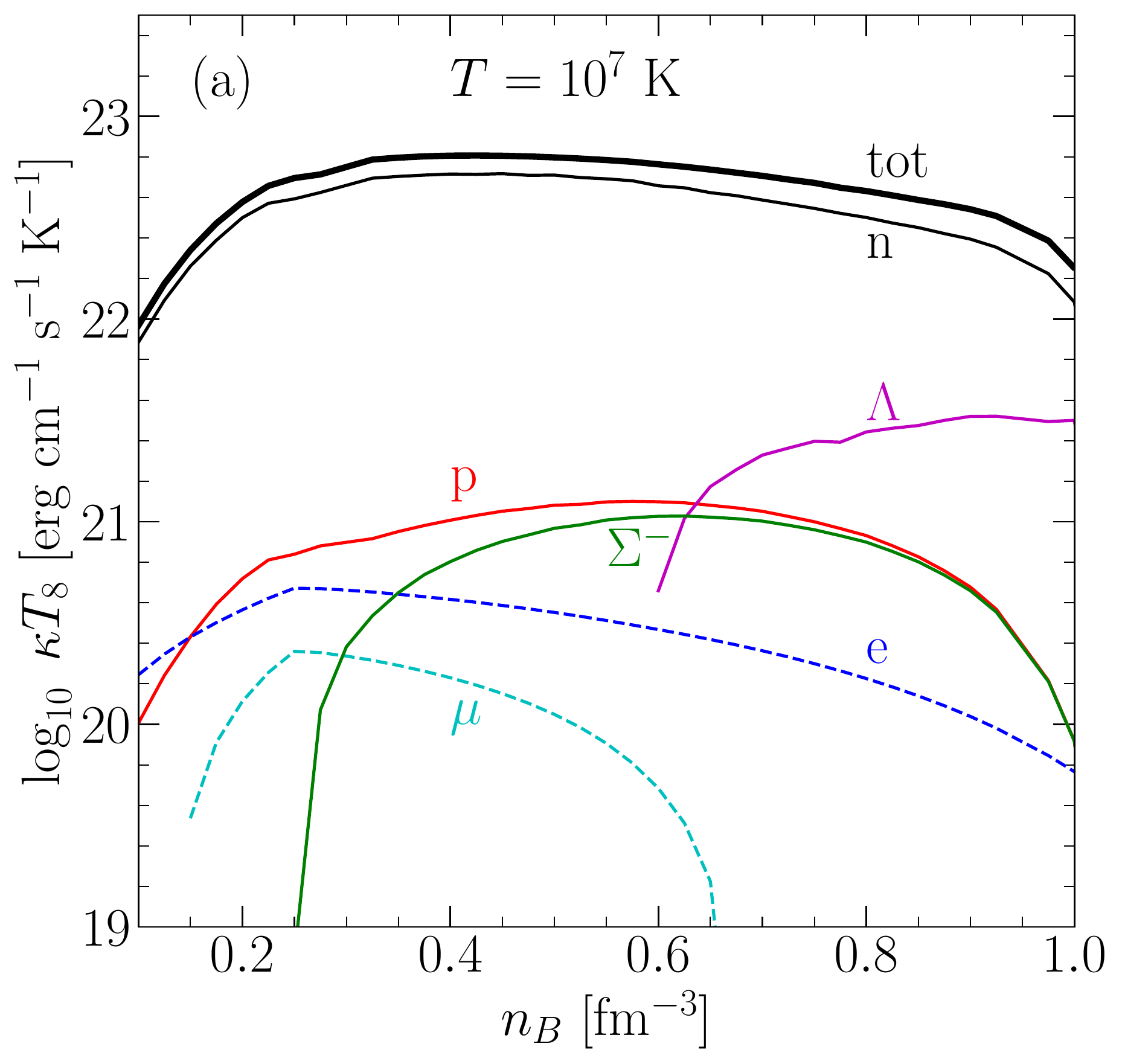}
\end{minipage}
\begin{minipage}{0.32\textwidth}
\includegraphics[width=\textwidth]{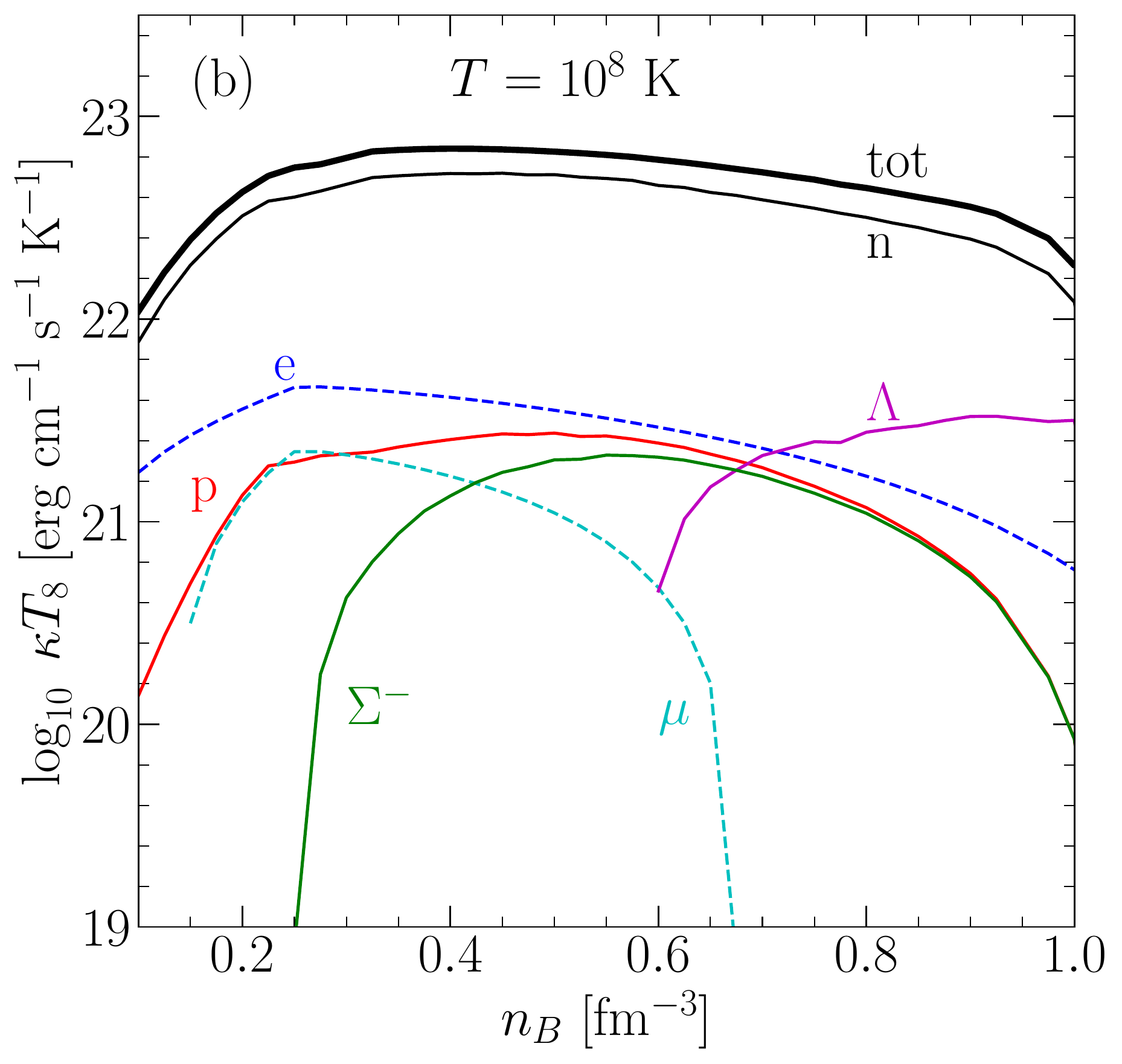}
\end{minipage}
\begin{minipage}{0.32\textwidth}
\includegraphics[width=\textwidth]{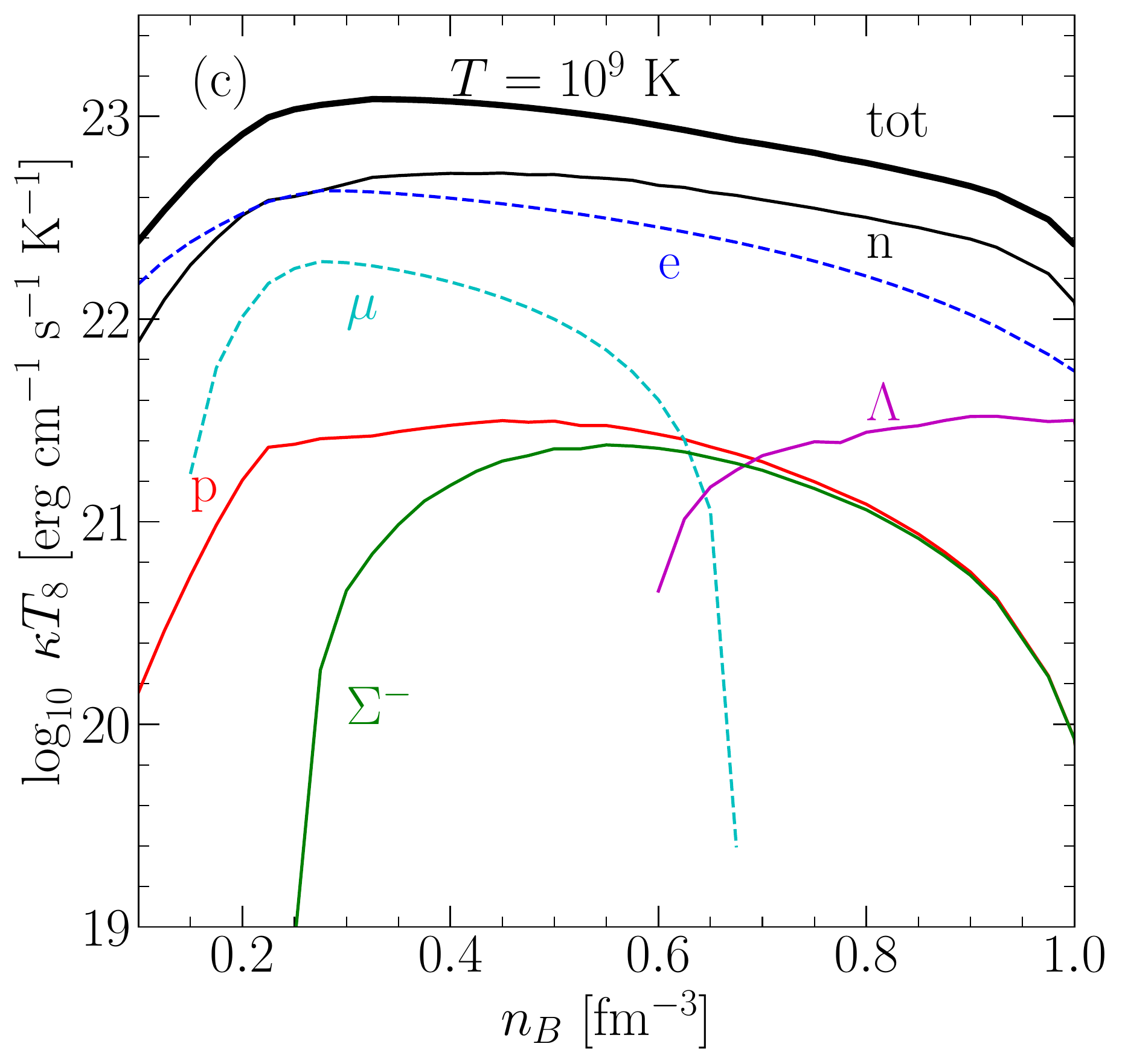}
\end{minipage}
\caption{Partial contributions to the thermal conductivity coefficient (variational solution) for the $\beta$-stable matter in NS cores. \new{Results for the NSC97e model are shown.}
Particle species labels are shown near the corresponding curves. The total thermal conductivity including corrections to the variational solution is shown with the thick solid line.
Results for three values of the temperature are shown: $T=10^7$~K \textbf{(a)},  $T=10^8$~K \textbf{(b)} and $T=10^9$~K \textbf{(c)}. 
 }\label{fig:kappa_part}
\end{figure}

\begin{figure}[H]
\widefigure
\begin{minipage}{0.32\textwidth}
\includegraphics[width=\textwidth]{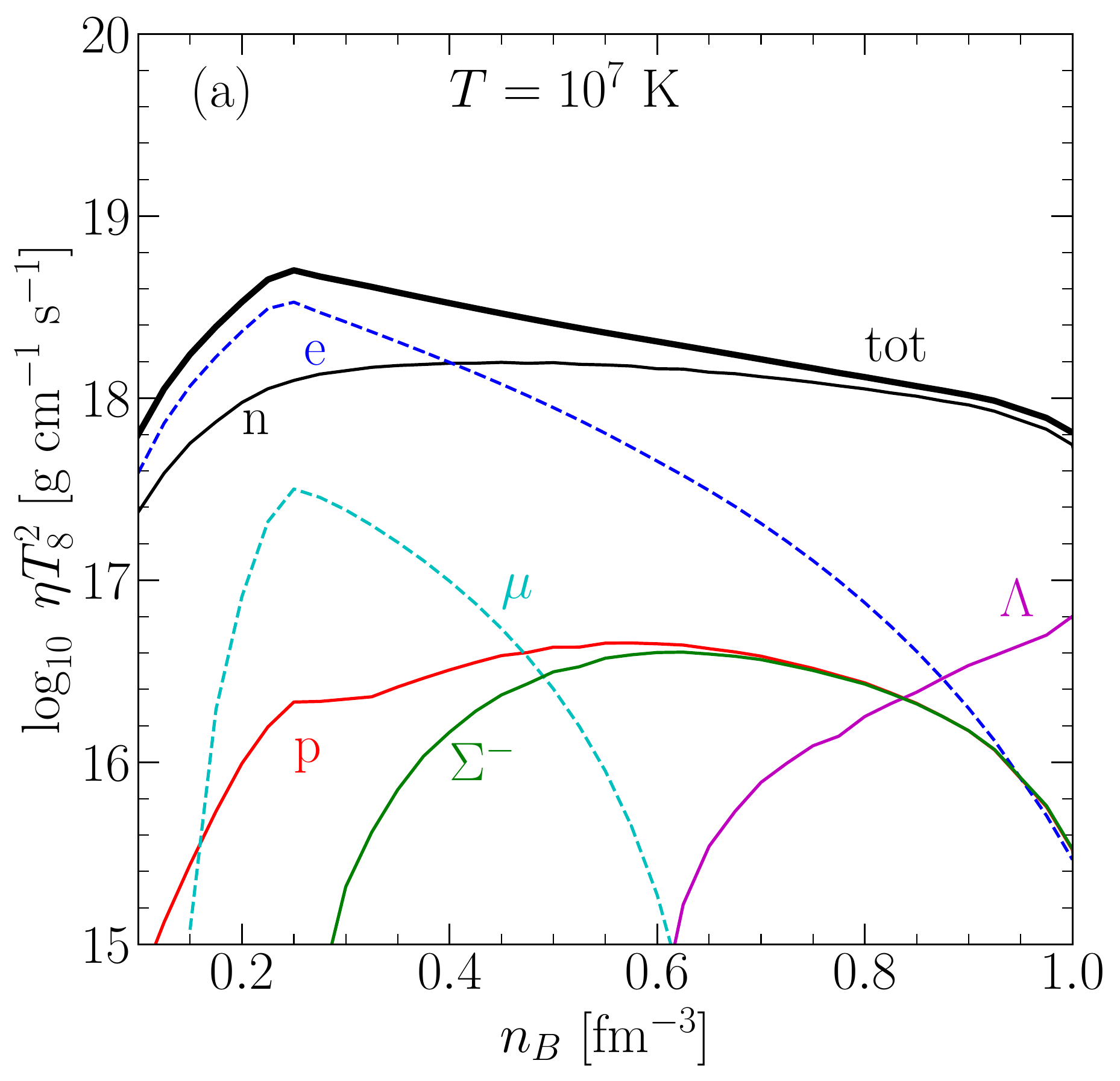}
\end{minipage}
\begin{minipage}{0.32\textwidth}
\includegraphics[width=\textwidth]{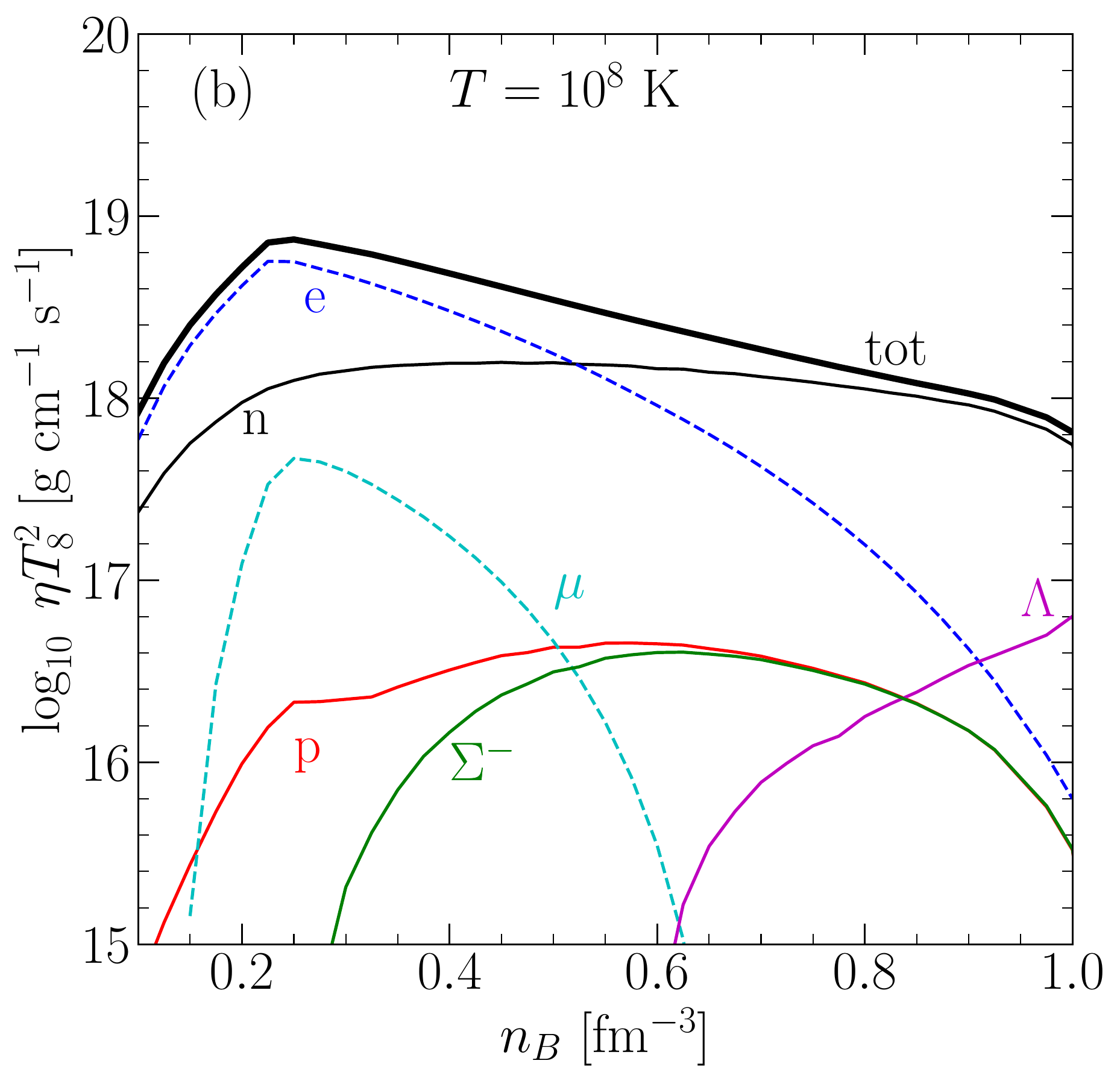}
\end{minipage}
\begin{minipage}{0.32\textwidth}
\includegraphics[width=\textwidth]{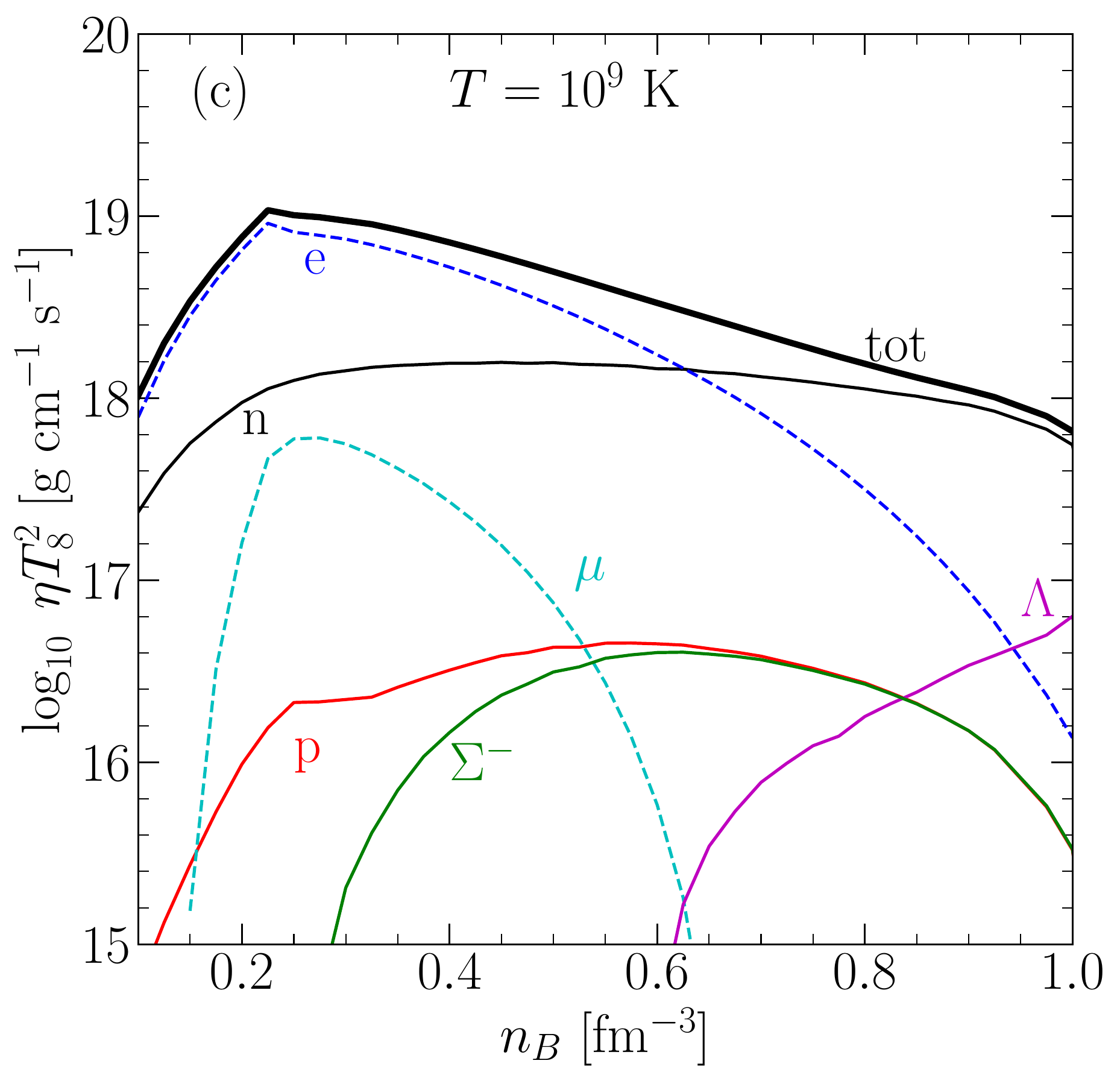}
\end{minipage}
\caption{Partial contributions to the shear viscosity coefficient (variational solution) for the beta-stable mater in NS cores. \new{Results for the NSC97e model are shown.} Particle species labels are shown near the corresponding curves. The total shear viscosity including corrections to the variational solution is shown with the thick solid line.
Results for three values of the temperature are shown: $T=10^7$~K \textbf{(a)},  $T=10^8$~K \textbf{(b)} and $T=10^9$~K \textbf{(c)}. 
}\label{fig:eta_part}
\end{figure}
\begin{paracol}{2}
\switchcolumn

Figures \ref{fig:kappa_part} and \ref{fig:eta_part} show, respectively, the partial contributions of the matter constituents to the thermal conductivity and shear viscosity coefficients. These contributions are obtained via the simplest variational solution.  The results are shown for the temperature-independent (in a normal Fermi-liquid) combinations $\kappa T_8$ and $\eta T^2_8$. Due to the non-Fermi liquid behavior of the lepton transport coefficients, we show results for three temperature values, $T=10^7,\, 10^8,$ and $10^9$~K. 
Thick solid lines in Figures~\ref{fig:kappa_part} and \ref{fig:eta_part} show total thermal conductivity and shear viscosity coefficients, respectively, including the corrections described in Section~\ref{sec:exact}.

\end{paracol}
\begin{figure}[t]
\widefigure
\begin{minipage}{0.45\textwidth}
\includegraphics[width=\columnwidth]{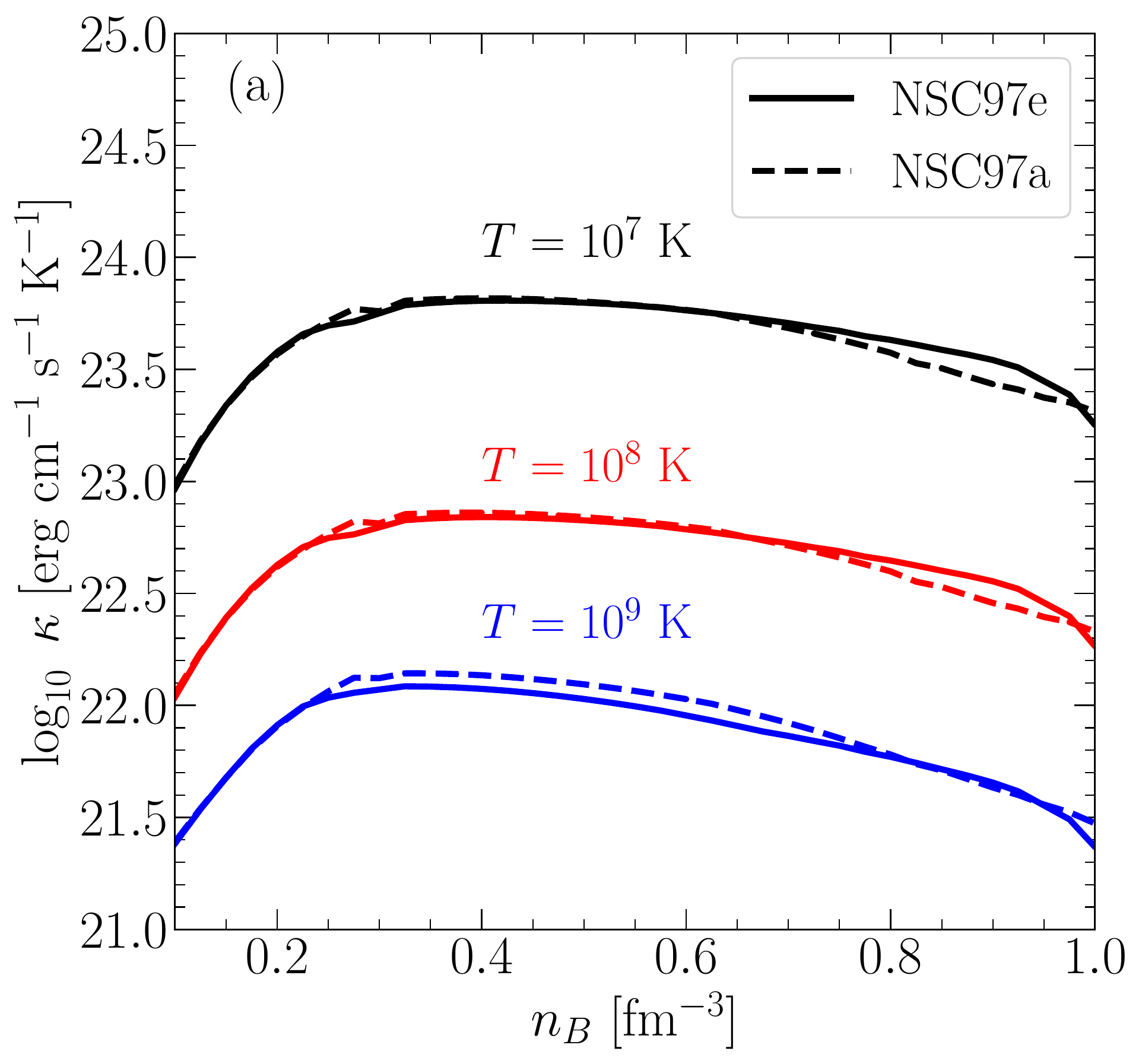}
\end{minipage}
\begin{minipage}{0.44\textwidth}
\includegraphics[width=\columnwidth]{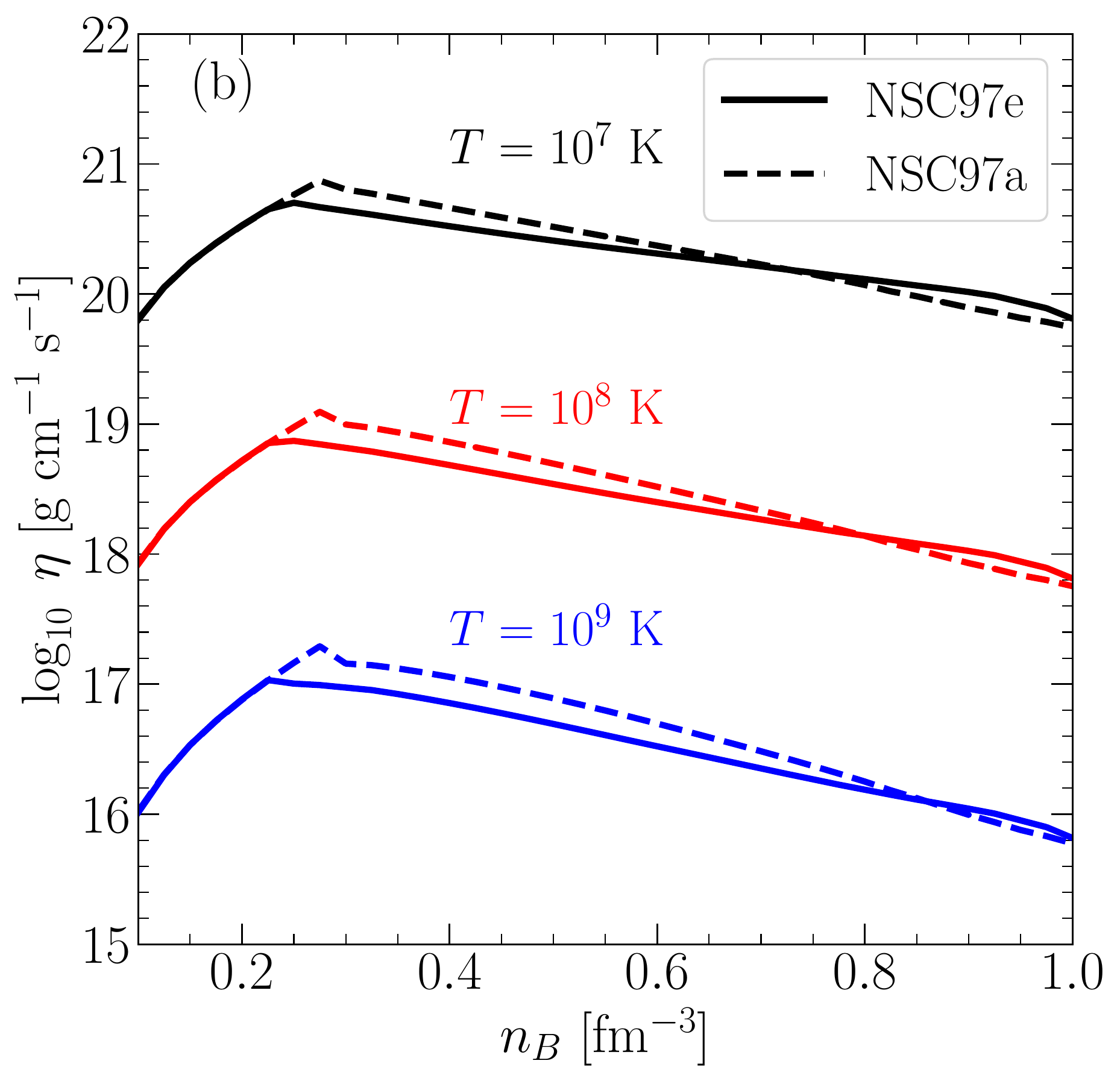}
\end{minipage}
\caption{\new{Total thermal conductivity \textbf{(a)} and shear viscosity \textbf{(b)} in hypernuclear NS cores for the NSC97e (solid lines) and NSC97 (dashed lines) models. Here we do not perform temperature rescaling of the transport coefficients. The results are shown for three values of temperature, $T=10^7$, $10^8$, and $10^9$~K as indicated in the plot.}} \label{fig:kappa_eta_comp}
\end{figure}
\begin{paracol}{2}
\switchcolumn

Among the baryon sector, neutrons always dominate the transport, like in the case of nuclear matter \cite{Shternin2013PhRvC,Shternin2020PhRvD}. In the case of the thermal transport, neutrons dominate also the total value of the thermal conductivity except for $T=10^9$~K where the lepton contribution becomes comparable to the nucleon one, see Figure~\ref{fig:kappa_part}c. This behavior is similar to \new{that} found in the nucleonic NS cores \cite{Shternin2013PhRvC,Shternin2020PhRvD}. 

A different situation is observed in the case of the shear viscosity, see Figure~\ref{fig:eta_part}. Here the leptonic contribution is dominant at lower densities (the upper boundary of this region depends on temperature), while neutrons give the main contribution at higher densities due to decrease in the lepton number fractions induced by the onset of the $\Sigma^-$ hyperon. This is in contrast with the results of the nucleonic matter where the lepton contribution was found to be always dominant in the non-superfluid matter \cite{Shternin2013PhRvC,Shternin2020PhRvD}. 
The total baryon contribution to $\eta$ is similar for both nucleon and hyperon NS core compositions. However, as a result of the suppression of lepton contribution, the total shear viscosity of the hypernuclear matter can be several orders of magnitude smaller than those in the nuclear matter.

\new{
Let us investigate the model dependence of the obtained results. Comparing the results for the  NSC9e and NSC97a potentials we find that although the partial transport coefficients exhibit some variations, the total values dominated by neutrons and leptons do not change much. This is illustrated in Figure~\ref{fig:kappa_eta_comp}, where we show thermal conductivity (panel (a)) and shear viscosity (panel (b)) for the NSC97e (solid lines) and NSC97a (dashed lines) for three values of temperature as indicated in the plot. Notice that here the absolute values of $\kappa$ and $\eta$ (without temperature rescaling) are plotted for convenience. The difference between the two models is not large. The reason is that the NSC97e and NSC97a models for the NY and YY interactions are quite similar and differ only on the values of few meson-baryon coupling constants, cut-off values of the vertex form factors, and some other parameters. Both describe, as mentioned before, with the same accuracy the existing hypernuclear data. In some sense this resembles the results for the nuclear matter of Ref.\ \cite{Shternin2020PhRvD}, where it was found that the main source for difference between the various models originates from the model for the three-body interactions which we do not include here for hyperons and do not alter for nucleons. Neutrons, as the most significant particles in terms of transport properties, are not affected by the change of model. The situation is different for other particles, see below.
}
\end{paracol}
\begin{figure}[H]
\widefigure
\begin{minipage}{0.45\textwidth}
\includegraphics[width=\columnwidth]{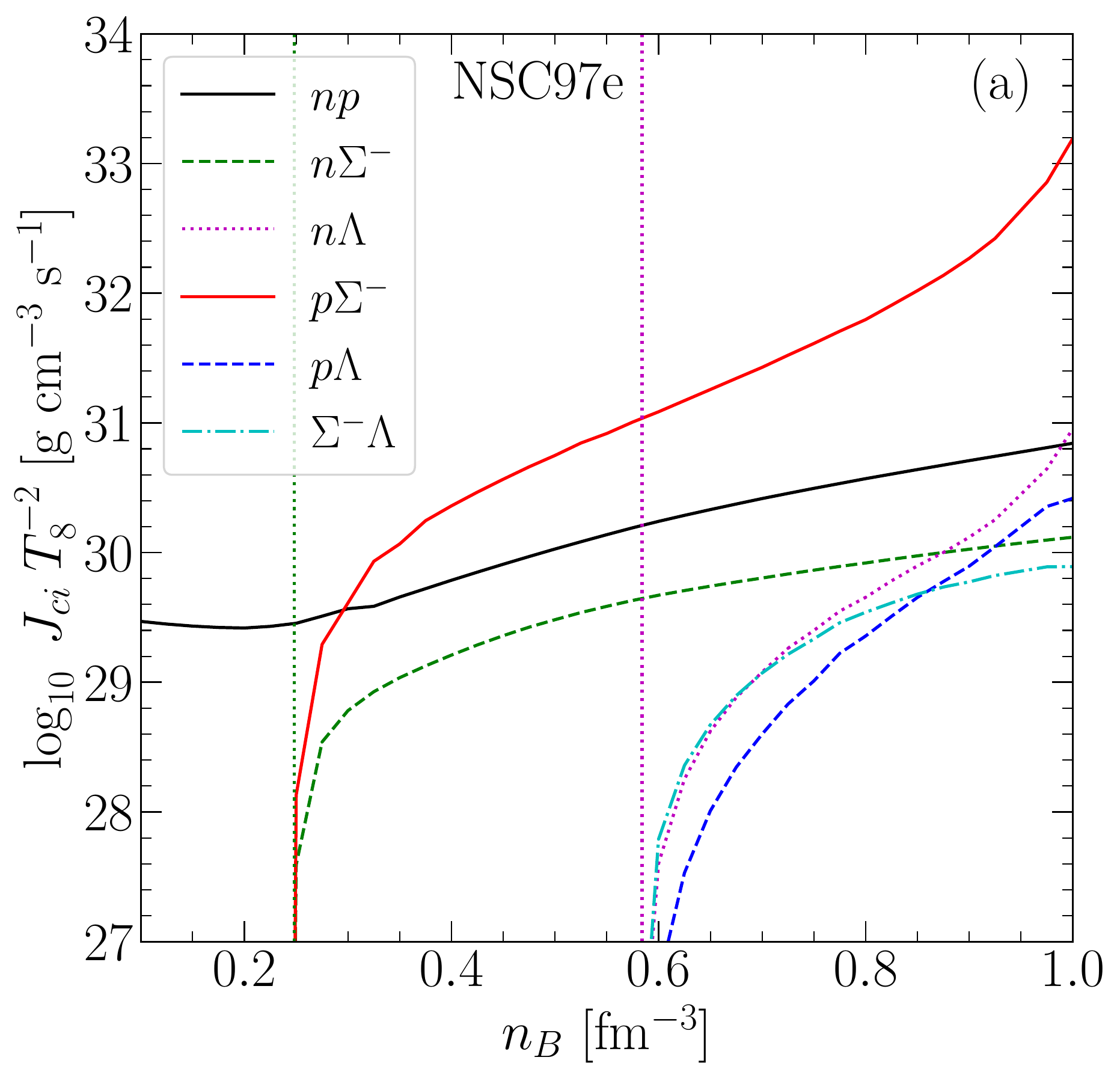}
\end{minipage}
\begin{minipage}{0.45\textwidth}
\includegraphics[width=\columnwidth]{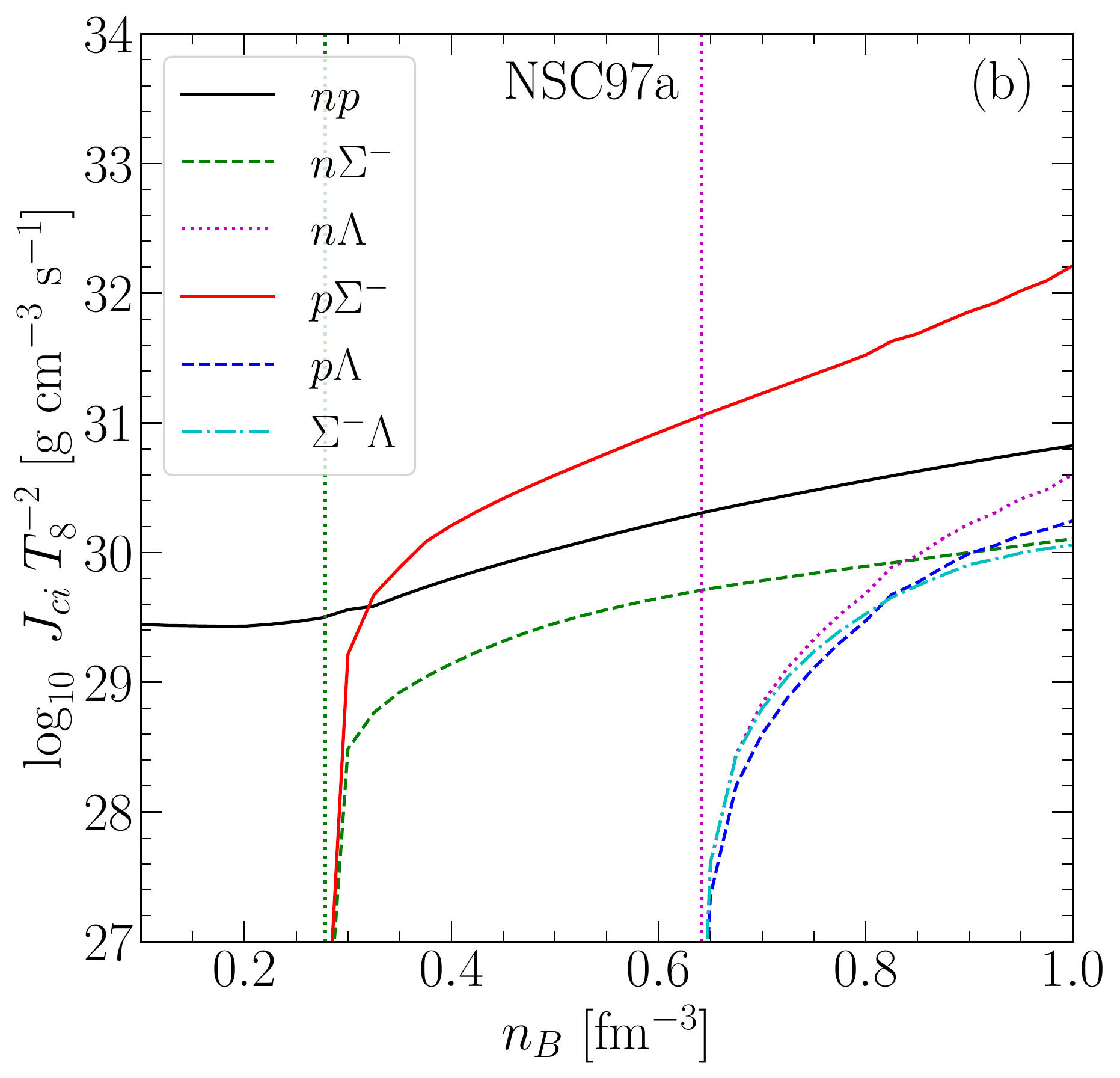}
\end{minipage}
\caption{Momentum transfer rates $J_{ci}$ mediated by the strong interactions \new{for the NSC97e (panel a) and NSC97a (panel b) models}. The  temperature-independent combination $J_{ci}T^{-2}_8$ is shown. Different lines correspond to different baryon pairs as shown in the legend. \new{Vertical dashed lines indicate $\Sigma^{-}$ and $\Lambda$ thresholds.}} \label{fig:Jci}
\end{figure}
\begin{paracol}{2}
\switchcolumn

Finally, in Figure~\ref{fig:Jci} we show the momentum transfer rates in the baryon collisions. \new{Panel (a) corresponds to the NSC97e model and panel (b) to the NSC97a model.}
Here, \new{we plot the temperature-independent combination 
$J_{ci}T_8^{-2}$.}
The contribution of the \new{electromagnetic interactions to $J_{ci}$ is negligible like in the case of the shear viscosity  above. }
The momentum transfer rates mediated by the electromagnetic interactions can be calculated following the expressions in Refs.\ \cite{Shternin2008JETP,Dommes2020PhRvD}. They are several orders of magnitude smaller than $J_{ci}$ mediated by the strong interaction, and we do not show them here. The momentum transfer rate between the nucleons, $J_{np}$, is in an order-of-magnitude agreement with the results of the pure nucleonic matter \cite{Shternin2020PhRvD}. A large value of $J_{p\Sigma^{-}}$ draws attention in Figure~\ref{fig:Jci}. This is due to the particularly strong interaction in this channel predicted by the \new{NSC97 models}, see Refs.\ \cite{Stoks1999PhRvC,Rijken1999PhRvC}. \new{Notice that for this transport coefficient the difference between the NSC97e and NSC97a model becomes apparent.}

In order to facilitate the application of our \new{study, we give}
the numerical results for the total thermal conductivity and shear viscosity coefficients, as well as the momentum transfer rates 
\new{in Tables~\ref{tab:kin} and \ref{tab:kin_NSC97a} in Appendix~\ref{app}.}

\section{Conclusions}\label{sec:conclude}

In this work, we have calculated the transport coefficients of the hypernuclear matter in the NS cores. In particular, we have calculated the thermal conductivity, the shear viscosity, and the momentum transfer rates for np$\Sigma^{-}\Lambda e\mu$ matter in $\beta$-equilibrium for baryon number densities between 0.1 and 1 fm$^{-3}$. The hypernuclear matter EOS was described within the non-relativistic BHF approach using realistic NN, NY, and YY interactions. Our results have shown that the proton, $\Sigma^-$, and $\Lambda$ do not give sizable contributions to the overall transport coefficients, being the 
total baryon contribution dominated by the neutron one as in the case of neutron star cores with only nucleons. However, these species are important in mediating the neutron mean free path.
In particular, we have found that neutrons dominate also the total thermal conductivity over the whole range of densities explored and that, due to the onset of $\Sigma^-$ which leads to the deleptonization of the neutron star core, they dominate also the shear viscosity on the high density region, in contrast with the pure nucleonic case where the lepton contribution is always the dominant one.

\new{Our paper makes only a first step in the studies of the transport properties of hypernuclear NS cores with realistic baryon interactions. The present study can be improved and extended in various aspects.

In particular, we employed  particular potentials both for nucleon and strange sectors and a specific many-body theory. Based on the experience for the nuclear sector \cite{Shternin2020PhRvD}, we expect that the variations of baryon interaction can lead to a substantial modifications of the values of the transport coefficients at large (e.g., $\gtrsim 2n_0$) baryon densities, especially when the three-body interactions are altered. Nevertheless, the qualitative conclusions presented above should be stable over the choice of a particular strong interaction model, which is seen comparing NSC97e and NSC97a results.}

\new{Next, the neutron star cores can contain strong magnetic fields. In this case the transport becomes anisotropic due to the Larmor rotation of the charged particles. The inclusion of these effects in not too strong magnetic fields, when the transport cross-sections are not modified, is relatively straightforward. For instance, the momentum transfer rates as calculated here constitute the microphysical input for the magnetic field evolution studies, see, e.g., Ref.\ \cite{Goldreich1992ApJ,Dommes2020PhRvD}. Large fields, e.g., $\gtrsim 10^{14}$~G, may affect the polarization tensor of the plasma making the scattering of the charged particles due to electromagnetic interactions anisotropic. This case is more complicated and requires a separate study.  In even stronger fields, e.g., $\gtrsim 10^{18}$~G, the particle motion becomes strongly quantized and the description of the transport based on the semiclassical distribution functions becomes invalid, however the presence of such ultra-high magnetic fields can hardly be considered as realistic.}

\new{Finally, due to the presence of the attractive  character  of  some partial wave channels of the baryon interaction, the baryon can be subject to pairing instabilities, see,  e.g., Ref.\ \cite{SedrakianClark2019EPJA}, for a review. Naturally, the strength and the character of the particular partial wave interaction is density-dependent.
It is expected that the protons in NS cores pair in the ${}^1$S$_0$ channel, while the neutrons should pair in the ${}^3$P$_2-{}^3$F$_2$ channel. Hyperons may also form superfluids if their interactions are attractive enough. However, a quantitative estimation of the hyperon pairing, contrary to the nucleonic one, has not received so much attention and just few calculations exist in the literature \cite{Balberg98,Takatsuka99,Takatsuka00,Takatsuka02,Vidana04,Zhou05,
Wang10}. In the presence of the paired species, the dynamical processes in NSs are described by a complicated multifluid hydrodynamics of the superfluid/suprerconducting mixtures e.g., \cite{Andersson2021Univ}, and references therein. Baryon pairing affects the transport coefficients of the NS core matter in several ways (see, e.g., \cite{Schmitt2018}, for review). We briefly outline the main effects expected in the presence of pairing. In this case the transport of the normal component of matter is governed by the Bogoliubov quasiparticles, which spectra have a gap at the Fermi surface. This results in exponential suppression of the scattering rates of quasiparticles at low temperatures. In NS core context, this effect was considered in Ref.\ \cite{Baiko2001AA} for neutron thermal conductivity and in Refs.\ \cite{ShterninYakovlev2007,ShterninYakovlev2008} for lepton transport coefficients in presence of the proton pairing. 
Moreover, the collision integral for the Bogouluibov quasiparticles modifies and should include processes with non-conserved number of quasiparticles, e.g.~\cite{Vollhardt1990}. In addition, if charged species are paired, the electromagnetic polarization of this superconducting matter is modified which affects the scattering rate of all charged quasiparticles (even non-superconducting ones, such as leptons) \cite{Schmitt2018, Shternin2020PhRvD}.
At low temperatures, the gapless Goldstone modes, e.g. superfluid phonons, can contribute to the transport properties of matter, see, e.g.~\cite{Manuel2021Univ}, for a review.

Even more complicated is the case of the magnetized superfluid/superconducting matter, where -- if the type II superconductivity is realized -- the topological structures called Abrikosov vortices appear. The study of all  these effects is much desirable but is beyond the scope of the present paper.}

\vspace{6pt} 



\authorcontributions{Authors contributed equally to this work}

\funding{A part of this work was done during the PHAROS COST STSM \#CA16214-44662. The work was partially supported by the Foundation for the Advancement of Theoretical Physics and Mathematics ``BASIS'', grant 17-13-205-1.}

\dataavailability{The data presented in this study are available on request from the corresponding author} 

\acknowledgments{\new{We thank the anonymous referees for suggestions which helped us to improve the paper.}
P.S. Thanks the INFN Sez.~di Catania for hospitality. }

\conflictsofinterest{The authors declare no conflict of interest.}

\abbreviations{Abbreviations}{The following abbreviations are used in this manuscript:\\

\noindent 
\begin{tabular}{@{}ll}
NS & Neutron star\\
EOS & Equation of state\\
BHF & Brueckner-Hartree-Fock\\
Av18 & Argonne v18 \\
UIX & Urbana IX\\
NSC97 & Nijmegen Soft-Core 97
\end{tabular}}

\appendixtitles{no} 
\appendixstart
\appendix
\section{}\label{app:meff}
\new{
We fit the dependence of effective masses on baryon density via the following expression
\begin{equation}\label{eq:meff_ft}
\frac{m^*_{c}}{m_c} = \sum\limits_{k=0}^4 a_{ck} \left(\frac{n_B}{1~\mathrm{fm}^{-3}}\right)^k
\end{equation}
for $c=n,\,p,\,\Sigma^{-},$ and $\Lambda$. The fit coefficients $a_{ck}$ are given in Table~\ref{tab:meff} for NSC97e model and in Table~\ref{tab:meff_a} for NSC97a model. Notice that in the latter case we used third-order polynomials.
}

\begin{specialtable}[h] 
\caption{Coefficients of the polynomial fit (\ref{eq:meff_ft}) for NSC97e model. Last column gives the  density range where the expression (\ref{eq:meff_ft}) is valid. \label{tab:meff}}
\begin{tabular}{l rrrrr l}
 \toprule
 species & $a_0$& $a_1$& $a_2$& $a_3$ & $a_4$ & $n_B$ range \\ 
 \midrule
$n$ & $0.940$ & $-0.668$ & $0.678$ & $0.029$ & $-0.121$ & $0.10-1$~fm$^{-3}$  \\ 
$p$ & $0.933$ & $-2.343$ &  $6.504$  & $-6.472$ &  $2.316$ & $0.10-1$~fm$^{-3}$ \\ 
$\Sigma^{-}$ & $0.853$  & $0.0418$ & $-0.590$ &  $0.695$ & $-0.239$ & $0.25-1
$~fm$^{-3}$ \\
$\Lambda$ & $-4.256$ &  $19.554$ & $-28.78$ &  $19.27$&  $-5.077$ &  $0.60-1
$~fm$^{-3}$ \\
 \bottomrule
\end{tabular}
\end{specialtable}

\begin{specialtable}[h] 
\caption{Coefficients of the polynomial fit (\ref{eq:meff_ft}) for NSC97a model. Last column gives the  density range where the expression (\ref{eq:meff_ft}) is valid. Notice that the $a_4$ coefficient is absent. \label{tab:meff_a}}
\begin{tabular}{l rrrr l}
 \toprule
 species & $a_0$& $a_1$& $a_2$& $a_3$ &   $n_B$ range \\ 
 \midrule
$n$ & $0.934$ & $-0.591$ & $0.453$ & $0.029$ & $0.10-1$~fm$^{-3}$  \\ 
$p$ & $0.826$ & $-1.092$ &  $2.171$  & $-1.000$ &   $0.10-1$~fm$^{-3}$ \\ 
$\Sigma^{-}$ & $0.905$  & $-0.459$ & $0.401$ &  $-0.078$  & $0.30-1
$~fm$^{-3}$ \\
$\Lambda$ & $0.892$ &  $-3.916$ & $8.742$ &  $-4.844$&   $0.65-1
$~fm$^{-3}$ \\
 \bottomrule
\end{tabular}
\end{specialtable}

\section{}\label{app}
\new{Tables~\ref{tab:kin} and \ref{tab:kin_NSC97a} contain } the numerical values of the transport coefficients for the hypernuclear NS core composition discussed in the main text \new{for NSC97e and NSC97a models, respectively}. We show $\kappa T_8$ (columns 2--4) and $\eta T_8^2$ (columns 5--7) for three values of the temperature. Total (lepton plus baryon) values of the transport coefficients are given. In the baryon subsystem the corrections for the variational solution for $n$ and $\Lambda$ are included. We also tabulate the momentum transfer rates $J_{ci}T_8^{-2}$ in collisions of six possible distinguishable baryon pairs (columns 7--13). 

\end{paracol}

\begin{specialtable}[h] 
\widetable
\caption{Transport coefficients of the hypernuclear matter as a function of baryon density \new{for the NSC97e model}. \textbf{Columns 2--4}: total thermal conductivity for three values of temperature. \textbf{Columns 5--7}: total shear viscosity for three values of temperature. For $\eta$ and $\kappa$ exact solutions of transport equations in the sense described in the text are shown. \textbf{Columns 8--14}: momentum transfer rates $J_{ci}$. \label{tab:kin}}

\begin{tabular}{ccccccc|cccccc}
 \toprule
 $n_B$ & \multicolumn{3}{c}{ $\log_{10} \kappa T_8 - 22$} &\multicolumn{3}{c}{ $\log_{10}\eta T_8^2-18$}&  \multicolumn{6}{c}{$\log_{10} J_{ci} T^{-2}_8 - 29 $}\\\\
  fm$^{-3}$ & \multicolumn{3}{c}{ erg~cm$^{-1}$s$^{-1}$~K$^{-1}$} &\multicolumn{3}{c}{ g~cm$^{-1}$s$^{-1}$}&\multicolumn{6}{c}{g~cm$^{-3}$s$^{-1}$}\\\\
  & $10^7$~K & $10^8$~K & $10^9$~K & $10^7$~K & $10^8$~K & $10^9$~K &$np$& $n\Sigma^{-}$&$n\Lambda$&$p\Sigma^{-}$&$p\Lambda$&$\Sigma^{-}\Lambda$\\
 \midrule
 0.100 & $-$0.032 & 0.038 & 0.384 & $-$0.201 & $-$0.078 & 0.012 		&   0.469 &--&--&--&--&--\\
 0.125 & 0.175 & 0.230 & 0.537 & 0.051 & 0.192 & 0.302 		        &   0.449 &--&--&--&--&--\\
 0.150 & 0.340 & 0.392 & 0.678 & 0.24 & 0.401 & 0.532 		        &   0.433 &--&--&--&--&--\\
 0.175 & 0.472 & 0.523 & 0.807 & 0.392 & 0.570 & 0.719 		        &   0.423 &--&--&--&--&--\\
 0.200 & 0.578 & 0.627 & 0.912 & 0.528 & 0.719 & 0.883 		        &   0.418 &--&--&--&--&--\\
 0.225 & 0.657 & 0.707 & 0.995 & 0.651 & 0.854 & 1.032 		        &   0.431 &--&--&--&--&--\\
 0.250 & 0.696 & 0.748 & 1.035 & 0.702 & 0.871 & 1.004 		        &   0.455 & $-$1.407 &--& $-$0.869 &--&--\\
 0.275 & 0.714 & 0.763 & 1.056 & 0.668 & 0.845 & 0.993 		        &   0.510 & $-$0.461 &--& 0.292 &--&--\\
 0.300 & 0.750 & 0.795 & 1.071 & 0.639 & 0.817 & 0.974 		        &   0.567 & $-$0.219 &--& 0.611 &--&--\\
 0.325 & 0.787 & 0.827 & 1.085 & 0.611 & 0.789 & 0.955 		        &   0.586 & $-$0.073 &--& 0.933 &--&--\\
 0.350 & 0.796 & 0.834 & 1.084 & 0.580 & 0.755 & 0.924 		        &   0.657 & 0.036 &--& 1.067 &--&--\\
 0.375 & 0.803 & 0.839 & 1.080 & 0.550 & 0.720 & 0.891 		        &   0.722 & 0.126 &--& 1.246 &--&--\\
 0.400 & 0.806 & 0.841 & 1.074 & 0.521 & 0.684 & 0.855 		        &   0.787 & 0.210 &--& 1.360 &--&--\\
 0.425 & 0.807 & 0.840 & 1.065 & 0.493 & 0.648 & 0.817 		        &   0.849 & 0.288 &--& 1.466 &--&--\\
 0.450 & 0.806 & 0.837 & 1.055 & 0.465 & 0.612 & 0.777 		        &   0.910 & 0.359 &--& 1.565 &--&--\\
 0.475 & 0.803 & 0.832 & 1.043 & 0.437 & 0.575 & 0.736 		        &   0.970 & 0.425 &--& 1.661 &--&--\\
 0.500 & 0.798 & 0.826 & 1.029 & 0.410 & 0.539 & 0.694 		        &   1.028 & 0.483 &--& 1.749 &--&--\\
 0.525 & 0.792 & 0.819 & 1.013 & 0.384 & 0.503 & 0.651 		        &   1.083 & 0.537 &--& 1.845 &--&--\\
 0.550 & 0.785 & 0.810 & 0.996 & 0.359 & 0.468 & 0.608 		        &   1.138 & 0.586 &--& 1.918 &--&--\\
 0.575 & 0.776 & 0.800 & 0.977 & 0.335 & 0.433 & 0.565 		        &   1.191 & 0.631 &--& 2.006 &--&--\\
 0.600 & 0.764 & 0.786 & 0.955 & 0.311 & 0.399 & 0.522 		        &   1.241 & 0.671 & $-$1.398 & 2.086 & $-$2.312 & $-$1.210 \\
 0.625 & 0.752 & 0.773 & 0.933 & 0.287 & 0.366 & 0.480		        &   1.287 & 0.707 & $-$0.746 & 2.172 & $-$1.470 & $-$0.641 \\
 0.650 & 0.738 & 0.757 & 0.908 & 0.262 & 0.333 & 0.437 		        &   1.331 & 0.741 & $-$0.378 & 2.257 & $-$0.992 & $-$0.328 \\
 0.675 & 0.722 & 0.740 & 0.884 & 0.238 & 0.300 & 0.395 		        &   1.374 & 0.773 & $-$0.118 & 2.343 & $-$0.655 & $-$0.103 \\
 0.700 & 0.707 & 0.724 & 0.864 & 0.213 & 0.267 & 0.352 		        &   1.417 & 0.804 & 0.078 & 2.428 & $-$0.400 & 0.071 \\
 0.725 & 0.688 & 0.705 & 0.842 & 0.188 & 0.234 & 0.310 		        &   1.456 & 0.835 & 0.262 & 2.522 & $-$0.169 & 0.216 \\
 0.750 & 0.672 & 0.688 & 0.820 & 0.164 & 0.203 & 0.269 		        &   1.495 & 0.865 & 0.400 & 2.612 & 0.010 & 0.334 \\
 0.775 & 0.648 & 0.664 & 0.793 & 0.138 & 0.171 & 0.227 		        &   1.533 & 0.893 & 0.548 & 2.707 & 0.223 & 0.459 \\
 0.800 & 0.632 & 0.647 & 0.770 & 0.115 & 0.141 & 0.188 		        &   1.570 & 0.920 & 0.655 & 2.797 & 0.357 & 0.540 \\
 0.825 & 0.610 & 0.625 & 0.744 & 0.090 & 0.111 & 0.150 		        &   1.605 & 0.948 & 0.782 & 2.907 & 0.508 & 0.612 \\
 0.850 & 0.588 & 0.601 & 0.715 & 0.065 & 0.082 & 0.113 		        &   1.640 & 0.975 & 0.897 & 3.019 & 0.654 & 0.680 \\
 0.875 & 0.567 & 0.579 & 0.687 & 0.041 & 0.054 & 0.078 		        &   1.675 & 1.001 & 1.000 & 3.137 & 0.776 & 0.734 \\
 0.900 & 0.542 & 0.554 & 0.656 & 0.015 & 0.025 & 0.043 		        &   1.709 & 1.026 & 1.118 & 3.269 & 0.895 & 0.774 \\
 0.925 & 0.508 & 0.520 & 0.616 & $-$0.015 & $-$0.008 & 0.005 		&   1.742 & 1.050 & 1.253 & 3.422 & 1.045 & 0.823 \\
 0.950 & 0.448 & 0.459 & 0.554 & $-$0.062 & $-$0.057 & $-$0.047 		&   1.776 & 1.073 & 1.449 & 3.639 & 1.200 & 0.857 \\
 0.975 & 0.387 & 0.398 & 0.491 & $-$0.109 & $-$0.106 & $-$0.099 	& 	1.809 & 1.096 & 1.645 & 3.855 & 1.355 & 0.890 \\
 1.000 & 0.251 & 0.264 & 0.368 & $-$0.190 & $-$0.187 & $-$0.183         &   1.843 & 1.118 & 1.954 & 4.192 & 1.418 & 0.892\\
 \bottomrule
\end{tabular}
\end{specialtable}

\begin{specialtable}[h] 
\widetable
\caption{\new{Same as Table~\ref{tab:kin} but for the NSC97a model.} \label{tab:kin_NSC97a}}

\begin{tabular}{ccccccc|cccccc}
 \toprule
 $n_B$ & \multicolumn{3}{c}{ $\log_{10} \kappa T_8 - 22$} &\multicolumn{3}{c}{ $\log_{10}\eta T_8^2-18$}&  \multicolumn{6}{c}{$\log_{10} J_{ci} T^{-2}_8 - 29 $}\\\\
  fm$^{-3}$ & \multicolumn{3}{c}{ erg~cm$^{-1}$s$^{-1}$~K$^{-1}$} &\multicolumn{3}{c}{ g~cm$^{-1}$s$^{-1}$}&\multicolumn{6}{c}{g~cm$^{-3}$s$^{-1}$}\\\\
  & $10^7$~K & $10^8$~K & $10^9$~K & $10^7$~K & $10^8$~K & $10^9$~K &$np$& $n\Sigma^{-}$&$n\Lambda$&$p\Sigma^{-}$&$p\Lambda$&$\Sigma^{-}\Lambda$\\
 \midrule
 0.100 & $-0.015$ & 0.052 & 0.391 & $-0.193$ & $-0.07$ & 0.022 		&0.446 & -- & -- & -- & -- & -- \\
 0.125 & 0.182 & 0.237 & 0.540 & 0.053 & 0.195 & 0.306 		    	&0.438 & -- & -- & -- & -- & -- \\
 0.150 & 0.338 & 0.39 & 0.677 & 0.239 & 0.401 & 0.532 		    	&0.435 & -- & -- & -- & -- & -- \\
 0.175 & 0.465 & 0.517 & 0.804 & 0.39 & 0.567 & 0.716 				&0.432 & -- & -- & -- & -- & -- \\
 0.200 & 0.568 & 0.619 & 0.907 & 0.525 & 0.716 & 0.88 	        	&0.432 & -- & -- & -- & -- & -- \\
 0.225 & 0.648 & 0.698 & 0.991 & 0.648 & 0.851 & 1.028 				&0.447 & -- & -- & -- & -- & -- \\
 0.250 & 0.715 & 0.766 & 1.062 & 0.764 & 0.977 & 1.164 		    	&0.467 & -- & -- & -- & -- & -- \\
 0.275 & 0.770 & 0.821 & 1.123 & 0.871 & 1.094 & 1.292 		    	&0.496 & -- & -- & -- & -- & -- \\
 0.300 & 0.759 & 0.813 & 1.122 & 0.803 & 0.996 & 1.158 	        	&0.559 & $-0.513$ & -- & 0.215 & -- & -- \\
 0.325 & 0.806 & 0.854 & 1.143 & 0.772 & 0.971 & 1.146 		    	&0.587 & $-0.235$ & -- & 0.673 & -- & -- \\
 0.350 & 0.812 & 0.858 & 1.143 & 0.735 & 0.937 & 1.121 		    	&0.664 & $-0.077$ & -- & 0.886 & -- & -- \\
 0.375 & 0.816 & 0.860 & 1.140 & 0.699 & 0.900 & 1.090 		    	&0.733 & 0.042 & -- & 1.082 & -- & -- \\
 0.400 & 0.818 & 0.860 & 1.135 & 0.662 & 0.861 & 1.056 		    	&0.798 & 0.144 & -- & 1.208 & -- & -- \\
 0.425 & 0.817 & 0.858 & 1.127 & 0.626 & 0.821 & 1.018 		    	&0.860  & 0.236 & -- & 1.318 & -- & -- \\
 0.450 & 0.814 & 0.854 & 1.118 & 0.589 & 0.780 & 0.978 		    	&0.918 & 0.318 & -- & 1.417 & -- & -- \\
 0.475 & 0.809 & 0.848 & 1.107 & 0.553 & 0.738 & 0.936 		    	&0.973 & 0.390 & -- & 1.509 & -- & -- \\
 0.500 & 0.803 & 0.841 & 1.094 & 0.516 & 0.696 & 0.891 		    	&1.027 & 0.454 & -- & 1.597 & -- & -- \\
 0.525 & 0.795 & 0.832 & 1.080 & 0.480 & 0.652 & 0.845 				&1.079 & 0.510 & -- & 1.681 & -- & -- \\
 0.550 & 0.787 & 0.822 & 1.064 & 0.444 & 0.608 & 0.797 		    	&1.130  & 0.506 & -- & 1.764 & -- &-- \\
 0.575 & 0.777 & 0.811 & 1.047 & 0.408 & 0.563 & 0.748 		     	&1.179 & 0.606 & -- & 1.846 & -- & -- \\
 0.600 & 0.765 & 0.798 & 1.028 & 0.372 & 0.518 & 0.697 			    &1.227 & 0.648 & -- & 1.925 & -- & --\\
 0.625 & 0.753 & 0.784 & 1.007 & 0.336 & 0.472 & 0.645 			 	&1.275 & 0.687 & -- & 2.003 & -- & -- \\
 0.650 & 0.729 & 0.760 & 0.979 & 0.301 & 0.427 & 0.591 				&1.320  & 0.723 & $-1.405$ & 2.079 & $-1.646$ & $-1.392$ \\
 0.675 & 0.707 & 0.737 & 0.951 & 0.264 & 0.381 & 0.538 				&1.362 & 0.754 & $-0.551$ & 2.153 & $-0.797$ & $-0.563$ \\
 0.700 & 0.685 & 0.714 & 0.922 & 0.227 & 0.335 & 0.484 	        	&1.402 & 0.784 & $-0.161$ & 2.227 & $-0.400$ & $-0.202$ \\
 0.725 & 0.660 & 0.688 & 0.889 & 0.190 & 0.288 & 0.428 				&1.442 & 0.812 & 0.115 & 2.300 & $-0.114$ & 0.045 \\
 0.750 & 0.634 & 0.660 & 0.854 & 0.151 & 0.240 & 0.371 				&1.480  & 0.841 & 0.332 & 2.375 & 0.112 & 0.237 \\
 0.775 & 0.605 & 0.630 & 0.816 & 0.111 & 0.191 & 0.312 			 	&1.518 & 0.868 & 0.520 & 2.447 & 0.305 & 0.394 \\
 0.800 & 0.574 & 0.598 & 0.781 & 0.070 & 0.140 & 0.251 				&1.555 & 0.895 & 0.684 & 2.523 & 0.472 & 0.526 \\
 0.825 & 0.528 & 0.552 & 0.739 & 0.020 & 0.083 & 0.184 			 	&1.591 & 0.921 & 0.887 & 2.629 & 0.675 & 0.654 \\
 0.850 & 0.505 & 0.529 & 0.710 & $-0.017$ & 0.035 & 0.124 			&1.627 & 0.948 & 0.979 & 2.686 & 0.770 & 0.745 \\
 0.875 & 0.469 & 0.492 & 0.672 & $-0.062$ & $-0.018$ & 0.059 		&1.662 & 0.974 & 1.110 & 2.773 & 0.888 & 0.828 \\
 0.900 & 0.434 & 0.457 & 0.633 & $-0.105$ & $-0.069$ &$-0.004$ 		&1.696 & 1.001 & 1.221 & 2.857 & 0.993 & 0.908 \\
 0.925 & 0.410 & 0.432 & 0.600 & $-0.142$ & $-0.113$ & $-0.061$ 	&1.730  & 1.027 & 1.307 & 2.925 & 1.054 & 0.949 \\
 0.950 & 0.374 & 0.395 & 0.557 & $-0.184$ & $-0.162$ & $-0.121$ 	&1.762 & 1.053 & 1.416 & 3.019 & 1.135 & 0.997 \\
 0.975 & 0.352 & 0.372 & 0.523 & $-0.215$ & $-0.199$ & $-0.169$		&1.794 & 1.080 & 1.485 & 3.097 & 1.179 & 1.032 \\
 1.000 & 0.311 & 0.330 & 0.473 & $-0.257$& $-0.246$ & $-0.224$      &1.824 & 1.106 & 1.603 & 3.213 & 1.244 & 1.060\\
 \bottomrule
\end{tabular}
\end{specialtable}

\begin{paracol}{2}
\switchcolumn
\end{paracol}
\clearpage
\reftitle{References}


\externalbibliography{yes}
\def\apj{Astrophys. J.}
\def\apjl{Astrophys. J. Lett.}
\def\apjs{Astroph. J. Suppl. Ser.}
\def\mnras{Mon. Not. R. Astron. Soc.}
\def\aap{Astron. Astrophys.}
\def\prc{Phys. Rev. {\rm C}}
\def\prb{Phys. Rev. {\rm B}}
\def\prd{Phys. Rev. {\rm D}}
\def\plb{Physics Letters {\rm B}}
\def\apss{Astroph. Space Sci.}
\def\pla{Phys. Lett.  A}
\def\ssr{Space Sci. Rev.}
\def\araa{Ann. Rev. Astron. Astrophys.}
\def\aj{Astron. J.}
\def\jphys{J. Phys.}
\def\npa{Nucl. Phys. A}
\def\nphysa{Nucl. Phys. A}
\def\npb{Nucl. Phys.  B}
\def\ijmpe{Int. J. Mod. Phys. E}
\def\ijmpd{Int. J. Mod. Phys. D}
\def\ijmpa{Int. J. Mod. Phys. A}
\def\sovast{Soviet Astronomy}

\end{document}